\documentclass[twoside,twocolumn,9pt]{article}
\usepackage{extsizes}
\usepackage[super,sort&compress,comma]{natbib} 
\usepackage[version=3]{mhchem}
\usepackage[left=1.5cm, right=1.5cm, top=1.785cm, bottom=2.0cm]{geometry}
\usepackage{balance}
\usepackage{mathptmx}
\usepackage{sectsty}
\usepackage{graphicx} 
\usepackage{lastpage}
\usepackage[format=plain,justification=justified,singlelinecheck=false,font={stretch=1.125,small,sf},labelfont=bf,labelsep=space]{caption}
\usepackage{float}
\usepackage{fancyhdr}
\usepackage{fnpos}
\usepackage{mathrsfs}
\usepackage{physics}
\usepackage[english]{babel}
\addto{\captionsenglish}{%
  
}
\usepackage{array}
\usepackage{droidsans}
\usepackage{charter}
\usepackage[T1]{fontenc}
\usepackage[usenames,dvipsnames]{xcolor}
\usepackage{setspace}
\usepackage[compact]{titlesec}
\usepackage{hyperref}
\title{A comprehensive study of complex non-adiabatic exciton dynamics in MoSi$_2$N$_4$}%Article title goes here instead of the text "This is the title"

\author{Mingran Kong$^{a}$, Shuichi Murakami$^{b,c}$ and Tiantian Zhang$^{b,c}$}
\date{%
    $^a$Fudan University, Yangpu District, Shanghai 200437, People{'}s Republic of China\\%
    $^b$Department of Physics, Tokyo Institute of Technology, Ookayama, Meguro-ku, Tokyo 152-8551, Japan\\%
    $^c$Tokodai Institute for Element Strategy, Tokyo Institute of Technology, Nagatsuta, Midori-ku, Yokohama, Kanagawa 226-8503, Japan
    \today
}

\begin{document}
\maketitle

%\author{Mingran Kong,\textit{$^{a}$}  Shuichi Murakami\textit{$^{b,c}$} and Tiantian Zhang,\textit{$^{b,c\dag}$}} %Author names go here instead of "Full name", etc.

%%%END OF TITLE AND AUTHORS%%%

%%%FONT SETUP - please do not change any commands within this section

%%%FOOTNOTES%%%

% \sffamily{\textbf{The abstract should be a single paragraph which summarises the content of the article. Any references in the abstract should be written out in full \textit{e.g.} [Surname \textit{et al., Journal Title}, 2000, \textbf{35}, 3523].}}\\%The abstrast goes here instead of the text "The abstract should be..."

\sffamily{\textbf{Excitons, which are composite boson quasi-particles composed of bound electrons and holes, have many fascinating properties and great potential in practical applications. 
Though experimental studies on exciton dynamics are well-developed, the $ab\ initio$ simulation ones still remain vacant until two years ago.
Here, we apply the density functional theory (DFT) and many-body perturbation theory (MBPT) on 2D MoSi$_2$N$_4$ to study its exciton-related physics and non-adiabatic ultrafast exciton dynamics theoretically and numerically for the first time.
Due to its wide band gap, large exciton binding energy and similar 2D hexagonal crystal structure {to} transition-metal dichalcogenides~(TMDs), we expect MoSi$_2$N$_4$ to have distinguished excitonic properties.
We calculate the photoluminescence~(PL) spectra with final states {as} bright excitons, yet lots of them are contributed by the dark ones, and the results match the experimental ones perfectly. 
We also study the dark-exciton-involved processes, which were barely studied in the past but dominate in many physical processes, and obtain several main results like: (i) High scattering rates over the whole Brillouin zone~(BZ) within the order of magnitude from 10$^{-2}$ fs$^{-1}$ to 10$^1$ fs$^{-1}$; (ii) Thorough analysis for the dynamics of the dark excitons at $\Lambda$ valley, which have negative effective mass and the highest scattering rate among several exciton states; 
(iii) Simulate the time-resolved evolution of the excitons after photo-excitation with the real-time Boltzmann transport equation~(rt-BTE) techniques, in which process excitons at $K/K^{\prime}$ valley play an important role; 
(iv) Exciton dynamics with spin-valley locking at $K/K^{\prime}$ valley are also discussed here; 
(v) A new approach is proposed for modulating the non-adiabatic effects for excitons, accompanied by a chiral phonon absorption/emission, by tuning the chirality of the external circularly polarized light. 
All the results show that the 2D material MoSi$_2$N$_4$ is an ideal platform to study the exciton-involved physics and has great application value.}}\\%The abstrast goes here instead of the text "The abstract should be..."

\rmfamily %Please do not remove this line.

%
%\section*{New Concept}
%In the past decades, $ab$ $initio$ studies on exciton dynamics were restricted to the adiabatic level. 
%Here, with the development of algorithm for non-adiabatic exciton dynamics recently, we conduct a comprehensive study on 2D material MoSi$_2$N$_4$. 
%Beyond the previous research that only focused on the excitons at $\Gamma$, here we study the phonon-assisted exciton dynamics distributing over the whole BZ, including the photoluminescence and exciton-phonon (Ex-Ph) scattering rates. The former result matches the experimental ones well, while the scattering rates show strong Ex-Ph coupling effects due to the values are as high as {the ones} in h-BN. 
%Phonon-assistant invervalley/intravalley scattering process are also studied for the dark excitons with the highest scattering rates at $\Lambda$ valley and such {investigations} are barely reached in previous studies. 
%Time-resolved exciton dynamics shows a clear pattern of an intervalley scattering tunnel from $\Gamma$ to $K/K^{\prime}$ after photo-excitation. 
%Due to the large spin splitting of electronic bands at $K/K^{\prime}$ valley, different valley excitons with spin-valley locking feature can be selected by modulating the chirality of the incident circularly polarized light, accompanied with a chiral phonon emission/absorption. 
%

\section*{Introduction}
Excitons were initially discovered to be pivotal in the photo-excitation, while gradually, they were also found to be dominant in a wide variety of physical processes, such as Bose-Einstein condensation~\cite{Jinhua2020} (BEC), Mott transition~\cite{Guerci2019}, superfluid~\cite{Boning2011} and non-Abelian braiding~\cite{Wu2022}. 
All {these} properties make excitons an increasingly crucial part of the excited-state physics and receive growing applications in nano-structured optical devices, such as the photovoltaic cells
~\cite{Menke2013,Zhang2019,Ehrler2012,Hedley2013}, light-emitting diodes~\cite{Chen2021,Mesta2013,Hasan2022,Hofmann2012}, nanocrystal emitter~\cite{Gramlich2021,Rodina2016} and quantum dots~\cite{Zhang2022,Gupta2021}. 
Moreover, two kinds of exciton transistors have been already realized based on their diffusion effect~\cite{Unuchek2018} and valley polarization physics~\cite{zhang2022chiral,Ciarrocchi2019}, respectively. 
The great applied potential in excitons is once again on display for circumventing the demise of Moore's Law, and it is urgent to discover more materials with eminent exciton {dynamic} properties and implement {the} corresponding comprehensive studies for finding splendid candidates.

Since the pace of theoretical research on exciton dynamics is slightly behind the pace of experiments and applications, there is a growing demand for a comprehensive theoretical description of exciton dynamics, especially at the $ab$ $initio$ computational level, which can help us to understand the exciton dynamics in materials directly. 
Although $ab$ $initio$ approaches are well-established to predict exciton binding energies, optical transitions, and radiative lifetimes, those for exciton dynamics and non-adiabatic processes have just been established recently. 
Through continuous attempts, scientists have finally achieved great success in describing the vivid and precise picture of exciton dynamics in real materials through in-depth analysis of exciton-involved many-body interactions, such as exciton-phonon (Ex-Ph)~\cite{Marini2008,Cannuccia2019,Paleari2019,Chen2020,Zhang2021,Xiang2021,Antonius2022}, exciton-exciton~\cite{Linardy2020,Erkensten2021} and exciton-polaron~\cite{Efimkin2021} scattering processes, and exciton-{plasmon} fission~\cite{Steinhoff2017}. 
With the knowledge of these critical mechanisms, especially the methodology of non-adiabatic dynamics proposed in Ref.~\cite{Chen2020}, we can have a better understanding of the exciton-involved transport patterns and their corresponding experimental phenomena. It not only greatly narrows the gap between the theoretical/numerical studies and experiments/applications, but also accelerates the innovation of exciton devices and will push them to a new level. 

% In this paper, 
For the reason that 2D material MoSi$_2$N$_4$ has a similar crystal structure {to} hexagonal transition-metal dichalcogenides (TMDs)\cite{Ugeda2014,Chernikov2014}, one may expect they will preserve similar physical properties {as well}, such as remarkable ambient stability, electronic band structures and strongly {bound} electron-hole pairs. 
These similarities will make MoSi$_2$N$_4$ an ideal candidate to {possess} excellent exciton-involved physical properties.
{
    The non-adiabatic effects and the dynamics of the excitons remain intriguing but veiled for a long time. With our self-developed codes, we successfully simulate a series of phonon-involved exciton dynamic processes and find that these properties are extraordinary in MoSi$_2$N$_4$ for the first time.
}

{
    In this paper, $ab$ $initio$ approaches based on the density functional theory (DFT) and many-body perturbation theory (MBPT) are utilized to calculate the electron, phonon and exciton properties of $MoSi_2N_4$.
    We explore many exotic exciton dynamic processes in MoSi$_2$N$_4$ for the first time, such as the transport patterns of Ex-Ph scattering, which has a significant influence on various exciton-involved physical properties in most cases, photoluminescence (PL), scattering rates~(SRs), and most importantly, time-dependent exciton Boltzmann transport during the photoexcitation process, which can be observed by time-resolved angle-resolved photoemission spectroscopy (trARPES). 
    We also note that our PL calculations are in perfect agreement with the experimental results, and all the results show MoSi$_2$N$_4$ an ideal platform {for studying} the non-adiabatic exciton dynamics.
}

\section*{Results and Discussion}

\begin{figure*}[ht!]
\centering
\includegraphics[width=1\textwidth]{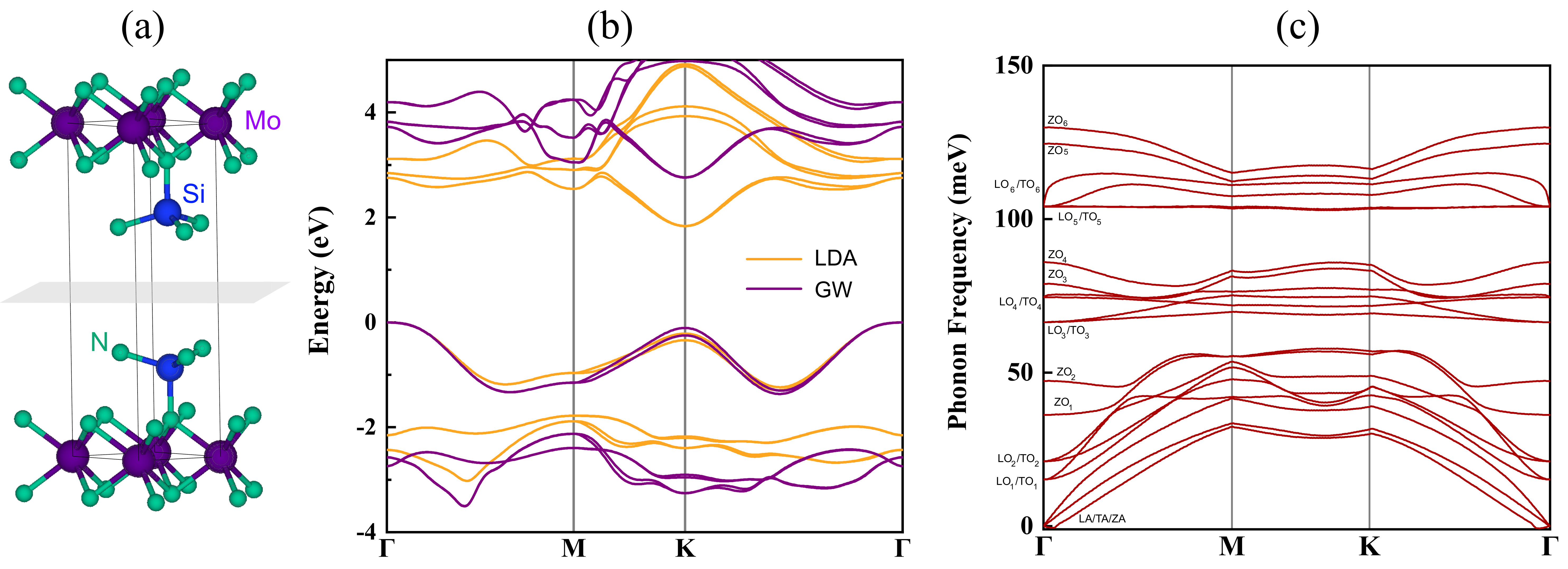}
\caption{(a) Crystal structure of MoSi$_2$N$_4$, where Mo, Si, N atoms are marked by the dark purple, blue and turquoise balls, separately.
(b) Spinful electronic structure of MoSi$_2$N$_4$ with the LDA method (orange lines) and the GW method (purple lines), respectively. 
The direct band gaps at $K$ are 2.00 eV (DFT) and 2.81 eV (GW), respectively. 
(c) Phonon dispersion for MoSi$_2$N$_4$.  
Vibration modes for each branch are labeled at $\Gamma$.}
\label{fig1}
\end{figure*}

\subsection{Crystal structure, electronic and phonon band structure for MoSi$_2$N$_4$}
MoSi$_2$N$_4$ and its family materials have attracted increasing interests recently {ever} since they {were} discovered, for having similar crystal structure with 2D hexagonal TMDs {and even} better physical properties~\cite{YiLun2020,Wu2021,Hong2020,Novoselov2020,li2020valley,chen2021first,Yao2021,yang2021accurate,wu2021semiconductor,zhong2021strain,Bafekry_2021,kang2021second,cui2021tuning,guo2020intrinsic,mortazavi2021exceptional,GUO2021110223,LI2021114753,wang2021article,cao2021two,wang2021intercalated,Yu2021,nandan2021two}. 
Figure~\ref{fig1} (a) is the crystal structure of MoSi$_2$N$_4$, which has the space group of \#187 and the point group $D_{3h}$. 
The crystal structure of MoSi$_2$N$_4$ can be treated as a MoN$_2$ prism layer sandwiched between two SiN$_4$ triangular pyramid layers, which are related by the $M_z$ symmetry marked by the gray plane. 
Due to the similarity in the crystal structure, MoSi$_2$N$_4$ also shares {a} similar electronic band structure with TMDs, as shown in Fig.~\ref{fig1} (b), i.e., there are two valleys located at $K$ ($\frac{1}{3}$, $\frac{1}{3}$) and $\Gamma$ (0, 0), which correspond to the conduction band minimum (CBM) and valence band maximum (VBM), respectively. 
Since the local density approximation~(LDA) approximation will always underestimate the band gap, the G$_0$W$_0$ method is also implemented to calculate the spinful band structure of MoSi$_2$N$_4$, as shown by the orange and purple lines in Fig.~\ref{fig1} (b). 
In this paper, all the calculations are based on the G$_0$W$_0$ method.
{
    The valence band maximum (VBM) is fixed at 0 eV in both LDA and GW calculations.
} 

Figure~\ref{fig1} (c) is the phonon dispersion of MoSi$_2$N$_4$, where the longitudinal-optical branch (LO) and transversal-optical branch~(TO) splittings vanish with the 2D Coulomb screening potential model~\cite{Sohier2017}. 
{Under the $D_{3h}$ point group, phonon modes for MoSi$_2$N$_4$ are composed of $\Gamma_{vib} = 4E'(I+R)\oplus 4A_2''(I) \oplus 3E''(R) \oplus 3A_1'(R) $ at $\Gamma$ point, where $I$ and $R$ represent the infrared and Raman activity, respectively. 
$E'$ and $E''$ correspond to the degenerate LA/TA and LO/TO branches, while $A_1'$ and $A_2''$ are the ZA and ZO branches, respectively.} 
Phonon frequencies {in} MoSi$_2$N$_4$ range from 0 meV to 129.88 meV, which offers a wide energy choice for the phonon-related scatterings. 
Thus, MoSi$_2$N$_4$ is an ideal platform {for studying} the Ex-Ph coupling, which plays an important role in the exciton dynamic processes.

\subsection{Excitons in MoSi$_2$N$_4$ and Ex-Ph scattering with final exciton states at $\Gamma$}

Excitons are quasiparticles composed of the electron-hole pairs bound by the Coulomb interactions. 
The effects of excitons are enhanced in 2D materials due to the reduced screening effects, which make{s} them ideal platforms for studying exciton-related physics. 
Since MoSi$_2$N$_4$ and TMDs share similar 2D hexagonal crystal structure and spinful electronic band structure, similar exciton spectra and exciton-involved processes are also expected in those two kinds of materials. 
Among the diverse exciton-involved processes, Ex-Ph coupling is always essential in {all} materials and has a great impact on all kinds of behaviors, such as the photoexcited process in TMDs. 
Thus, it's important to implement a study on these Ex-Ph scattering's impacts before focusing on the exciton dynamics. 

\begin{figure*}[htbp]
\centering
\includegraphics[width=\textwidth]{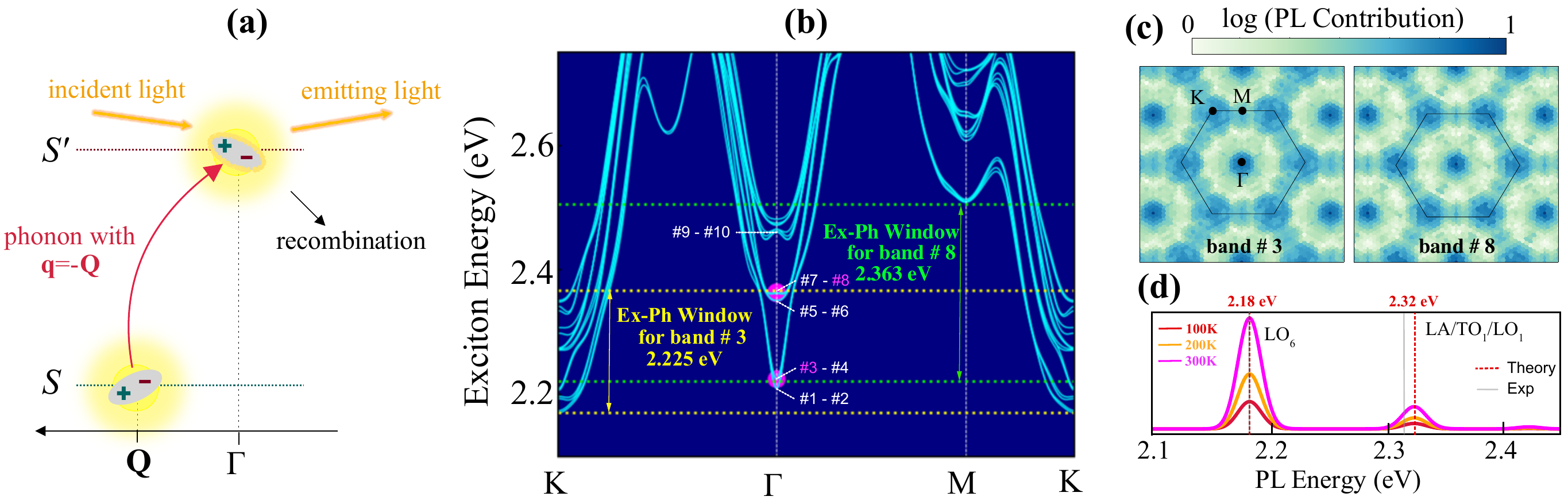}
\caption{(a) Illustration of the phonon-assisted photoluminescence process for excitons, where excitons with $\mathbf{Q}\neq0$ will be scattered to the bright exciton states at $\Gamma$. 
(b) Exciton spectra of the lowest 15 exciton bands along the high symmetry lines. 
The lowest 10 bands \#1-\#10 are marked at $\Gamma$, and the pink dots highlight two bright excitons states at \#3 (2.225 eV) and \#8 (2.363 eV) bands. 
The energy windows for Ex-Ph coupling of two bright excitons are labeled by yellow (\#3) and green (\#8) dotted lines, respectively.
(c) Momentum-resolved (for both exciton with $\mathbf{Q}$ and phonon with $\mathbf{q}=-\mathbf{Q}$) photoluminescence spectrum contribution in the exciton's momentum space (given in a log scale) by bright excitons of \#3 (left) and \#8 (right).
(d) Integrated photoluminescence spectrum for MoSi$_2$N$_4$ at 100 K, 200 K and 300 K marked by red, orange and pink lines, separately. 
Two peaks in our calculation (2.18 eV \& 2.32 eV) match with those two observed by experiments peaks marked by the gray lines with an energy shift.}
\label{fig2}
\end{figure*}

Although excitons are widely distributed in both momentum and frequency spaces, the ones receiving extensive attention are those with zero {momentum} ($\mathbf{Q}=0$ at $\Gamma$) and lower energies. 
The reason is not only because they are relatively easily excited, but also because only the excitons with $\mathbf{Q}=0$ can couple with photons.  
Excitons with $\mathbf{Q}=0$ and having significantly strong coupling strength with photons are called bright excitons. 
As shown in Fig.~\ref{fig2} (a), the bright exciton at state $S^{\prime}$ with $\mathbf{Q}=0$ can couple with an external photon, which will be emitted once the electron-hole pair recombines.
{Here, the phonon-assisted photoluminescence process can also be obtained, during which} the bright photon at state $S^{\prime}$ with $\mathbf{Q}=0$ can be supplied by the dark exciton at {its} initial state $S$ with $\mathbf{Q}\neq 0$ with the help of the phonon with $\mathbf{q}=-\mathbf{Q}$. 
{
    Therefore, excitons with non-zero momentum are receiving more and more attention in recent days.
}
The emitting lights encode the information of this whole process, which is documented in the PL spectrum and observable by experiments.

Figure~\ref{fig2} (b) is the exciton spectra of MoSi$_2$N$_4$, where two bright excitons with the lowest energies are marked by pink dots, with energies of 2.22 eV (\#3 exciton) and 2.36 eV (\#8 exciton), respectively. 
The exciton spectra shown in Fig.~\ref{fig2} (b) 
{
    are already modified with
}a red-shift of 0.33 eV to match the bright excitons at 2.21 eV and 2.35 eV obtained in experiments~\cite{Hong2020}. Such difference is probably caused by the 0 K approximation in our calculation and the room temperature in experimental observation. We note that the difference of energies may be reduced by implementing a calculation on a denser k-mesh, without changing the overall band configuration {of} excitons.  
However, since the Ex-Ph coupling calculations based on these spectra are enough to study the corresponding exciton dynamics, while the computing burden with a denser k-mesh is very heavy, we will study the exciton-involved physical processes based on the exciton spectra shown in Fig.~\ref{fig2} (b).

The energy windows for the Ex-Ph coupling of those two bright excitons are marked by yellow and green dashed lines in Fig.~\ref{fig2} (b), where the Ex-Ph scatterings are more likely to happen, based on the phonon frequency range of this material. 
Furthermore, within those two windows, one can notice that the main scattering processes for those two bright excitons can be classified into intravalley and intervalley ones, since the excitons at $\Gamma$ valley and $K/K^{\prime}$ valley have the lowest energies. 
The intravalley and intervalley exciton scatterings can be obtained with the help of phonons carrying momenta $\mathbf{q}=(0,0)$ and $\mathbf{q}$ = $K$ ($\frac{1}{3}$, $\frac{1}{3}$) / $K^{\prime}$ ($-\frac{1}{3}$, $-\frac{1}{3}$), respectively. %which habitated in $\Gamma$ and $K/K^{\prime}$ valleys, . 
It is worth {mentioning} that the intervalley Ex-Ph scattering processes can help to obtain bright excitons scattered from dark excitons at $\mathbf{Q}\neq0$ by phonons with $\mathbf{q}=-\mathbf{Q}$, i.e., Ex-Ph coupling can deliver and transfer the dark excitons with $\mathbf{Q}\neq0$ to the bright exciton states at $\Gamma$. 
Since only bright excitons can couple strongly with the light, such process of probing dark excitons out of $\Gamma$ is essential in many photo-exciton-involved processes. 

Therefore, in order to figure out which momentum (for both exciton with $\mathbf{Q}$ and phonon with $\mathbf{q}=-\mathbf{Q}$) dominates {in} the Ex-Ph scattering process and generates bright excitons with $\mathbf{Q}=0$, momentum-resolved phonon-assisted photoluminescence (PL) contribution pattern for MoSi$_2$N$_4$ with final exciton states of band \#3 and band \#8 (two bright excitons marked by the pink dots at $\Gamma$) are calculated by following Eq.~(\ref{PL}) and displayed in Fig.~\ref{fig2} (c). 
From the PL contribution patterns, we conclude that excitons at both $\Gamma$ and $K/K^{\prime}$ dominate in the Ex-Ph scattering process. 
The high intensity at $\Gamma$ is mainly contributed by the initial states as bright excitons with $\mathbf{Q}=0$, which is very common and widely discussed due to their strong coupling strength with photons, while the high intensity at $K/K^{\prime}$ is contributed by the initial states as dark excitons with $\mathbf{Q}\neq0$. 
Since the final states fall into bright exciton states at \#3 and \#8 in the PL process, the intervalley exciton scatterings have a higher possibility to happen between $K$ ($K^{\prime}$) and $\Gamma$ with phonon carrying $\mathbf{q}=-\mathbf{Q}=(\pm\frac{1}{3}, \pm\frac{1}{3})$ in MoSi$_2$N$_4$, contributing to the phonon-assisted PL process. 

In order to have a direct comparison with experimental results, we also calculated the integrated phonon-assisted photoluminescence (PL) spectra with Eq.~(\ref{PL}), which shows the information of bright (dark) excitons directly (obliquely) in MoSi$_2$N$_4$. 
As shown in Fig.~\ref{fig2} (d), two peaks with different energies and intensities correspond to those two bright excitons labeled in Fig.~\ref{fig2} (b). 
Energies for two bright excitons obtained by the PL spectra are 2.18 eV and 2.32 eV, respectively, which show a red-shift about 0.4 eV comparing with 
{
    their
} original exciton energies shown in Fig.~\ref{fig2} (b). Such difference is mainly induced by the Ex-Ph self-energy correction from the phonon branches of $6^{th}$ LO (LO$_6$) and longitudinal-acoustic/$1^{st}$ TO/$1^{st}$ LO branch (LA/TO$_1$/LO$_1$)~\cite{Chen2020}. 
We note that both the {intensities} and the energies obtained in our calculation match the experimental ones.
{ 
    As shown in Fig.~\ref{fig2} (d), the grey lines representing the experimental results have been already modified with a 0.17 eV blueshift
}\cite{Huang2022}, and such blueshift is common and usually caused by three reasons: 
(i) PL spectra in our calculation are on a 2D structure, while the experimental one is on a multilayer film, which will make an influence on the band gap; 
(ii) The one-shot G$_0$W$_0$ method in our calculation will also underestimate the band gap; 
(iii) Our calculation is performed on the electronic and exciton band structure calculated with the DFT and MBPT methods, {which fix the energy levels with the zero temperature approximation,} while the experimental one is at the room temperature, which will also make a difference on the band gap.
{
    Enlightened by the idea of applying external modulation such as strains and electric field to realize the manipulation of the exciton properties~\cite{Liang2022,Pak2017}, 
}
we {propose} that one can modulate the exciton {dynamic} process by tuning the exciton band structure via the same techniques, in which process information can be encoded with different phonon-assisted PL contribution patterns and diagnosed by the photon emitting from the bright excitons at $\Gamma$. 
{
    For example, the exciton state with the lowest energy at the $M$ point is on the edge of the upper Ex-Ph scattering window shown in Fig.~\ref{fig2}(b). 
    By applying strains or electric fields, one can modulate the energy of this state in or out the window and realize the manipulation of exciton dynamic properties. 
}

We also compare the PL spectrums at different temperatures, as shown in Fig.~\ref{fig2}~(d). 
For our calculation on MoSi$_2$N$_4$, the locations of the PL peaks are temperature independent while the intensity grows with the increase of the temperature. 
{Although our DFT and MBPT calculations are all based on the 0 K ground-state calculations and thus the eigenvalues of excitons won't change as the temperature increases, the occupation number will still rise up, leading to an enhancement of the intensity.}
Such phenomenon is attributed to the growth of the occupation number of both excitons and phonons as temperature increases, which also demonstrates the non-negligible contribution from dark excitons. 
Actually, except for the processes end with $\textbf{Q}=0$ exciton states mentioned here, dark excitons also dominate in various physical process{es}, especially the photo-excitation-excluded ones with final states out of the $\Gamma$ point.
Thus, thorough and detailed discussions on the dark excitons are of great significance and will begin from the next section.

\subsection{Ex-Ph scattering for MoSi$_2$N$_4$: excitons with initial and final states distributed in the whole BZ } 

\begin{figure*}
\centering
\includegraphics[width=\textwidth]{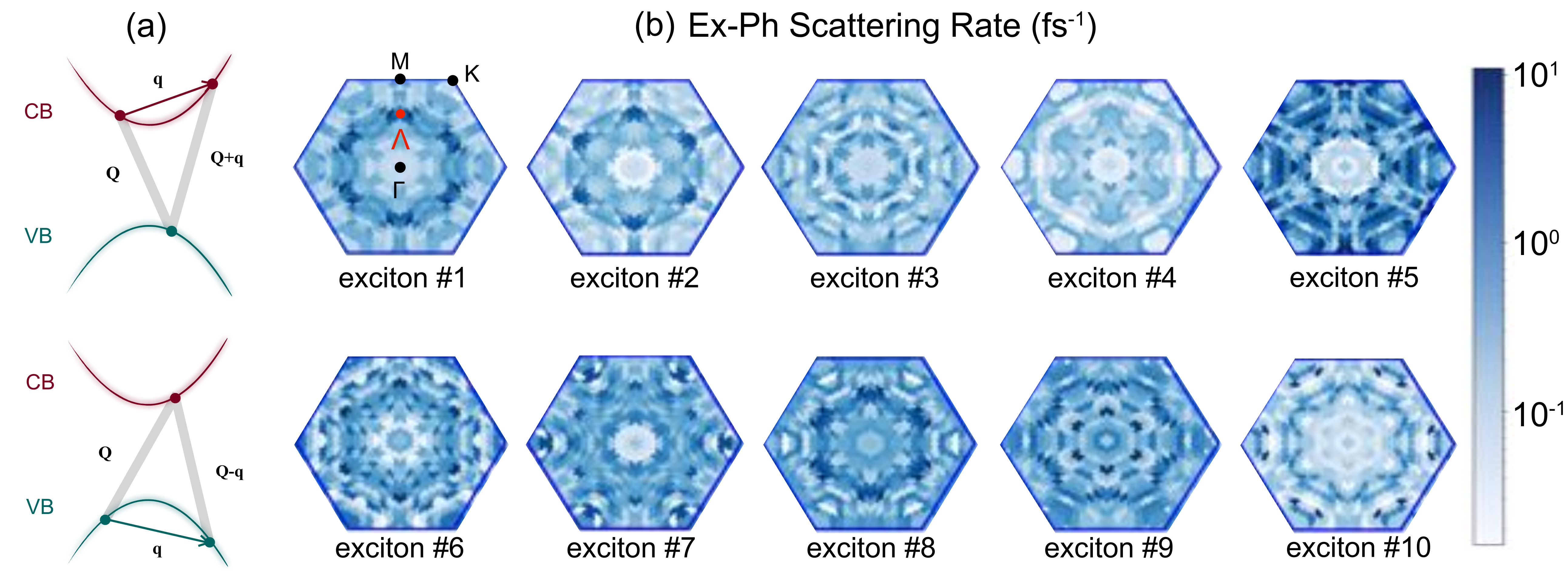}
\caption{(a) Illustration for two exciton scattering process, involving with phonon and exciton carrying momenta distributed over the whole BZ. 
The Ex-Ph scattering can be decomposed to the phonon-involved electron (top) and hole (bottom) scattering processes. 
(b) Ex-Ph scattering rates for MoSi$_2$N$_4$ in the whole BZ, where ten subfigures correspond to the lower ten exciton states. For exciton states \#1, \#2, \#6 and \#8-\#10, the scattering rates have a relatively large value at $\Lambda$. }
\label{fig3}
\end{figure*}

Driven by their exotic physical properties and the explosion of applications, studies on excitons have gradually stepped out the $\textbf{Q}=0$ {point} over the past five years, which opens the door to the long-forgotten physics of dark excitons with $\textbf{Q}\neq 0$.
Besides, Ex-Ph coupling process has remained {intriguing but} mysterious for a long time, especially for the dark ones, which have a wide distribution in the momentum space and make differences in all kinds of physical processes. 
In this paper, we study the Ex-Ph scattering rates in the whole BZ where dark excitons {dominate}, based on the Ex-Ph coupling scenarios described in Eq.~(\ref{ExPhMatrix}) and Fig.~\ref{fig3}~(a). 
Since excitons can be projected to different electron-hole pairs, the phonon-involved exciton scattering process can be treated as the electron~(hole) scattering process by phonons at conduction~(valence) band{s}. 
For example, the {bound} electron within an exciton carrying momentum $\textbf{Q}$ in the conduction band will be scattered to a new state with the help of phonon carrying $\textbf{q}$, which will transit {this} electron-hole pair to a new exciton state with momentum $\textbf{Q}+\textbf{q}$. 
A similar process can be obtained for {a} {bound} hole, yet with a final exciton state carrying momentum $\textbf{Q}-\textbf{q}$. 
With the help of Ex-Ph coupling scenarios and the self-energy relaxation time approximation (SERTA), Ex-Ph scattering rates at arbitrary {momentum} containing almost all the phonon-induced dumping effects can be obtained~\cite{Antonius2022}.

Figure~\ref{fig3}~(b) illustrates the Ex-Ph scattering rates for the lowest 10 exciton bands, showing {a} scale range{s} from $10^{-2}$ fs$^{-1}$ to $10^1$ fs$^{-1}$, which are as large as h-BN with {almost} the same scale about 0.01 fs$^{-1}$ - 20 fs$^{-1}$~\cite{Chen2020}. 
Since the Ex-Ph process is one of the most relevant factors for the lifetime of dark excitons, those figures show us how to exploit excitons with different momenta. 
For example, two exciton bands split by the spin-orbit coupling often have close energies, while their scattering rates may show a large difference, such as the ones for exciton \#1 and exciton \#2 in Fig.~\ref{fig3} (b). 
Furthermore, like TMDs, MoSi$_2$N$_4$ also has the characteristics of valley polarization, which can help to select the excitons with different Ex-Ph effects using corresponding external circularly polarized light, which will be discussed in section~$0.6$.

We also notice that for exciton branches \#1, \#2, \#6 and \#8-\#10, excitons at $\Lambda$ valleys (marked by the red dot in Fig.~\ref{fig3} (b)) have the largest scattering rates showing a much higher possibility to be scattered to other momenta/states, and thus we will implement a thorough study on them, {only the excitons at $\Lambda$ valley}, in the next section, especially for the intervalley scattering process{es}.

\begin{figure*}[ht!]
\centering
\includegraphics[width=\textwidth]{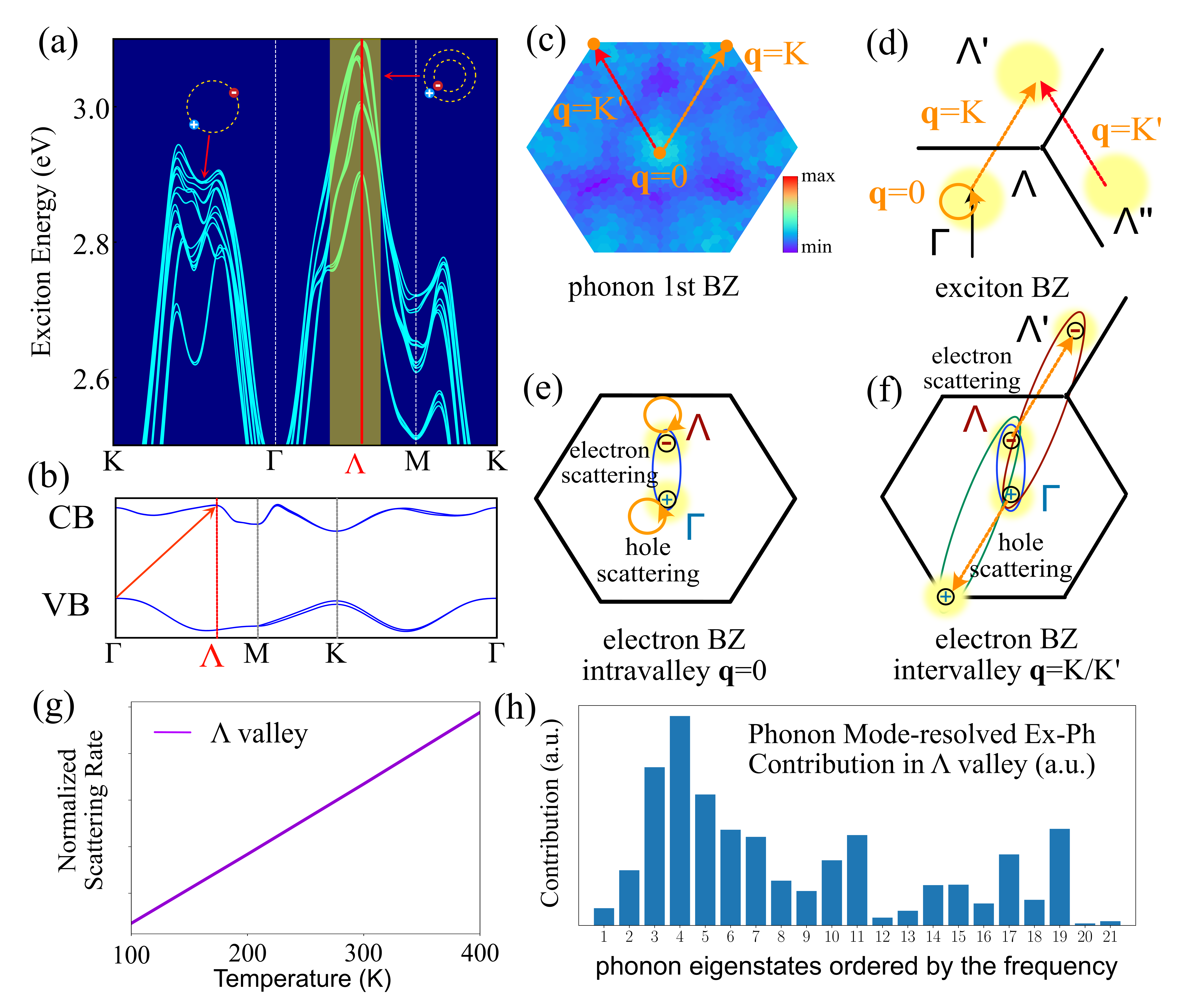}
\caption{ (a) Exciton spectra for the ones with higher energies, with $\Lambda$ valley marked by the red vertical line and orange shaded rectangle. 
(b) The main sources of the electron-hole pair are marked by the red arrow for the lowest two excitons at $\Lambda$. 
(c) $\mathbf{q}$-resolved strength map for the Ex-Ph coupling with lower 10 exciton states at $\Lambda$ valley.
(d) Illustration for the Ex-Ph intravalley ($\mathbf{q}$=0) scattering and intervalley ($\mathbf{q}$=$K$/$K'$) in the $\mathbf{Q}$ space. 
(e) Illustration for the Ex-Ph intravalley scattering with excitons at $\Lambda$ valley in $\mathbf{q}$ space, where two orange arrows correspond to electron scattering and hole scattering with $\mathbf{q}$ = 0. 
(f) Illustration for the Ex-Ph scattering between different $\Lambda$ valleys in $\mathbf{q}$ space, where two orange arrows correspond to electron scattering and hole scattering with $\mathbf{q}$ = $K$/$K'$ shown in Fig.~\ref{fig3} (a). 
(g) Normalized scattering rates for the lowest ten exciton states in the vicinity of $\Lambda$ valley under different temperatures, which shows a nearly linear correlation. 
(h) Phonon mode-resolved Ex-Ph contribution for the lowest ten excitons at $\Lambda$ valley.
}
\label{fig4}
\end{figure*}
%%%

\subsection{Ex-Ph scattering process with initial exciton states at $\Lambda$ valley}

{
    Starting from the simplified free exciton's Hamiltonian written as $H = \frac{\hbar^2\textbf{Q}^2}{2(m_e^* + m_h^*)} + \frac{\hbar^2\textbf{p}^2}{2m_\gamma^*}$, where $m_e^*$($m_h^*$) is the electron(hole)'s effective mass, $m_\gamma^*$ is the reduced mass, $\textbf{Q}$ is the momentum of the whole exciton and $\textbf{p}$ is the relative momentum describing the relative motion of the electron-hole pair to their orbit center. Therefore, the effective mass of exciton $\textbf{M}$ ($\textbf{M} = m_e^* + m_h^*$) is directly related to the second-other derivative of the exciton dispersion. Therefore we can obtain that
} 
the excitons at $\Lambda$ valleys have negative masses, which is not common and makes their corresponding Ex-Ph {dynamic} process {even more valuable to be discussed}, as shown in Fig.~\ref{fig4}~(a). 
{
    Excitons with negative mass are usually composed of the electron-hole pair joint orbiting around their center, i.e., their orbital center is on their same side, as shown in the right-inserted figure in Fig.~\ref{fig4}~(a), rather than the hydrogen-like one with positive mass as shown in the left-inserted figure. Therefore, the electron and hole forming exciton with negative mass will accelerate in the same direction under the external fields\cite{Lin2021}, which is quite counterintuitive and exotic for their exciton dynamic properties.
}
They are mainly contributed by the electrons with momentum $\mathbf{k}=\Lambda$ and holes with momentum $\mathbf{k}=\Gamma$. 
For example, the {lowest} two exciton states (\#1 and \#2) at $\Lambda$ are contributed by the electron-hole pair marked by the red arrow in Fig.~\ref{fig4}~(b). 
To have a better understanding of the strong Ex-Ph coupling at $\Lambda$ valley, phonon-momentum ($\mathbf{q}$)-reserved coupling strength of the lowest 10 exciton states is illustrated in Fig.~\ref{fig4}~(c).  
It shows that phonons with $\textbf{q}=0$ and $\textbf{q}=K/K^{\prime}$ contribute the most, corresponding to the intravalley and intervalley scattering for excitons with initial states at $\Lambda$. 
Since there is a large energy difference between the excitons at $\Lambda$ valley and other momenta, neither of the intravalley and intervalley scattering processes will make the exciton jump out of $\Lambda$ valley according to our analysis. 

In the $\Lambda$ intravalley scattering process, excitons transit between different exciton energy states with phonons carrying $\textbf{q}=0$, as marked by the orange circle at $\Lambda$ valley shown in Fig.~\ref{fig4}~(e). 
In the $\Lambda$ to $\Lambda^{\prime}$ intervalley scattering process corresponding to a transition between two neighboring BZs, phonons carrying $\textbf{q}=K$ are involved, as shown by the orange dashed line in Fig.~\ref{fig4}~(d). 
Similarly, a $\Lambda^{\prime\prime}$ to $\Lambda^{\prime}$ intravalley scattering process crossing two neighboring BZs can be also obtained with the assistance of phonons carrying $\textbf{q}=K^{\prime}$, as shown by the red dashed line in Fig.~\ref{fig4}~(d). 
Since excitons can be understood as collective coupling electron-hole pairs, the intervalley scattering process can be also understood as the electron scattering from $\Lambda$ to $\Lambda^{\prime}$ or hole scattering from $\Gamma$ to $K$, as shown by the orange dashed line in Fig.~\ref{fig4}~(f). 

We also study the temperature dependence of scattering rates for the excitons at $\Lambda$ valley from 100 K to 400 K, as shown in Fig.~\ref{fig4}~(g), which displays a distinct linear-like proportional pattern.
This is caused by the increase of the occupation numbers of both exciton{s} and phonon{s} as the temperature increases.
Figure~\ref{fig4}~(h) is the phonon-mode-resolved Ex-Ph contribution for excitons at $\Lambda$ valley, where phonon modes \#4, \#5 and \#6 contribute the most to the Ex-Ph coupling and phonon modes \#1, \#12, \#20 and \#21's are prohibited in the Ex-Ph processes. We note that most of the phonon modes are with $\textbf{q}=0$ or $\textbf{q}=K/K^{\prime}$, based on the Ex-Ph coupling strength for excitons with initial states at $\Lambda$ valley in Fig.~\ref{fig4}~(c).

\subsection{Time-dependent non-adiabatic exciton dynamics in MoSi$_2$N$_4$}
Compared with the time-resolved ultrafast dynamics experimental approaches, that have developed rapidly in recent years, such as tr-ARPES, the dynamic method being able to be used to study the time-resolved excitons from the viewpoint of first-principles has been barely realized. 
In order to have a vivid insight into the exciton non-adiabatic {dynamic} process, we employ rt-BTE here to simulate the exciton diffusion process after a strong photo-excitation from the semi-classic point of view. 
Such a process will start from the light-pulse-induced non-equilibrium state, where most excitons gather in the bright exciton states at $\Gamma$, and end with equilibrium ones. 
Exciton distributions in the $\mathbf{Q}$ space are calculated at the time of 20 fs, 100 fs and 200 fs, separately, as shown in Figure~\ref{fig5} (a). 
At 20 fs, almost all the excitons gather at $\Gamma$, as marked by the single red point, corresponding to the beginning of the non-equilibrium process after photo-excitation. 
As time goes by, a distinguished pattern is obtained at 100 fs, i.e., excitons evolve from $\Gamma$ to $K/K^{\prime}$, and it remains until to 200 fs. Afterward, the system returns to the equilibrium state, which also matches our Ex-Ph scattering rates calculation. 
Such a dynamic process can be observed by experiments like tr-ARPES, which will be discussed in our future work by combining theoretical and experimental results.

\subsection{Spin-valley exciton dynamics under polarized photo-excitation}
\begin{figure*}[ht!]
\centering
\includegraphics[width=\textwidth]{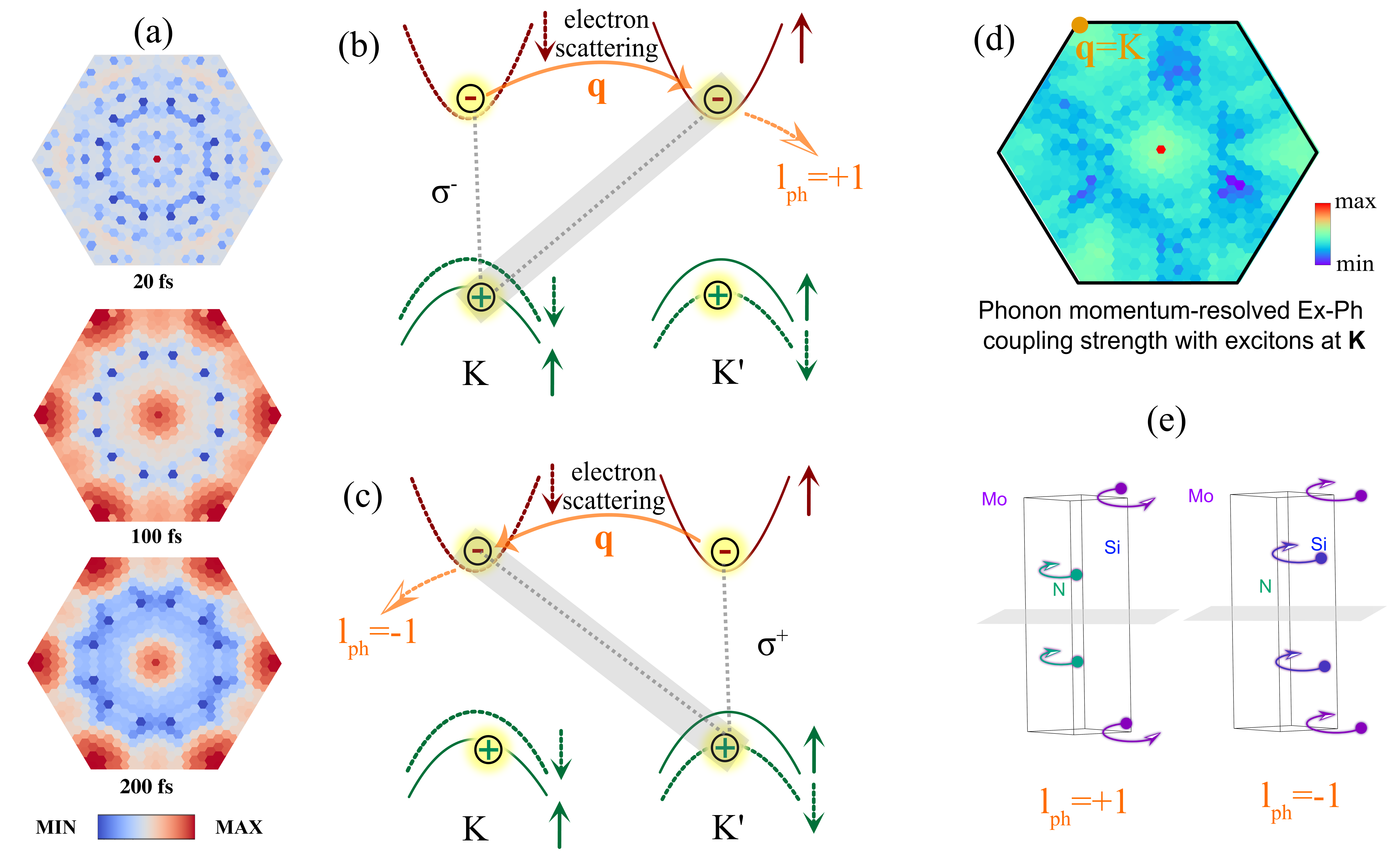}
\caption{(a) rt-BTE simulation of the non-adiabatic dynamics for the bright exciton \#3 at 20 fs, 100fs and 200fs after an external light impulse, where the occupation number is shown in log scale.
(b-c) Circularly polarized light-induced valley polarization for excitons through the electron-scattering process. 
(b) Under a left-hand circularly polarized light ($\sigma^{-}$), electrons in the valence band at $K$ with spin $\uparrow$ will be scattered to the conduction band at $K^{\prime}$, with the assistance of chiral phonon emitting carrying $\mathbf{q}$ = $K$ and $l_{ph}$ = +1. 
(c) Under a right-hand circularly polarized light ($\sigma^{+}$), electrons in the valence band at $K^{\prime}$ with spin $\downarrow$ will be scattered to the conduction band at $K$, with the assistance of chiral phonon emitting carrying $\mathbf{q}$ = $-K$ and $l_{ph}$ = $-1$. 
(d) $\mathbf{q}$-resolved Ex-Ph coupling strength for the lower two excitons at $K$, corresponding to the electron-hole bounds marked by the grey shaded rectangles in (b-c). It shows that phonon with $\mathbf{q}=K$ has a relatively strong coupling strength.
(e) Vibration modes of the two chiral phonons at $\mathbf{q}=K$ with $l_{ph}$ = $+1$, $-1$, respectively.
}
\label{fig5}
\end{figure*}

As shown in Fig.~\ref{fig5}~(a), the system will transit excitons from $\Gamma$ to $K/K^{\prime}$ after photo-excitation, in which process the $\Gamma\to$ $K/K^{\prime}$ intervalley scattering plays an important role. 
Thus, in this section, we will focus on the exciton dynamics located at $\Gamma$ and $K/K^{\prime}$ valleys, especially from the valley polarization point of view. 
In addition, since the irreducible representations~(irreps) for excitons still remain unclear, an electron-hole pair picture is introduced for the selection rule analysis. 
Therefore, excitons with $\mathbf{Q}=\Gamma$ and $\mathbf{Q}=\ K/K^{\prime}$ can be decomposed to the $K\to K$ (or $K^{\prime}\to K^{\prime}$) {bound} electron-hole pairs and $K\to K^{\prime}$ (or $K^{\prime}\to K$) {bound} pairs, respectively. 
The intervalley scattering{s} for exciton \#2 as electron-hole pairs are shown in Figs.~\ref{fig5}~(b-c). 
Since the spin splittings for the valence bands in MoSi$_2$N$_4$ are non-negligible, they will contribute to the photo-selective valley polarization together with a spin polarization in this material, resulting in spin-valley locking effects. Thus, different exciton states with different Ex-Ph effects can be alternatively selected, such as exciton \#1 and exciton \#2 with distinct scattering rates shown in Fig.~\ref{fig3} (b). 

By following the selective rules of spin, energy and momentum, the electron in the valence located at $K$ valley with the spin $\uparrow$ state will be excited to a virtual conduction state at $K$ by a left-hand circularly polarized light $\sigma^{-}$, and it will be further scattered to the $K^{\prime}$ valley with spin $\uparrow$, {via} the assistance of phonon carrying $\mathbf{q}=K$. 
We note that such a process also preserves pseudo-angular momentum ($l_{ph}$) conservation\cite{zhang2015chiral,Tiantian2022}, which will make the process accompanied by emitting (absorbing) a chiral phonon carrying $l_{ph}$ = +1 ($-1$). 
Chiral phonons with different $l_{ph}$ will have different vibration modes in the real space, which are illustrated in Fig.~\ref{fig5} (e).  
Likewise, by changing the external light from left-hand $\sigma^{-}$ to right-hand $\sigma^{+}$, electrons located at $K^{\prime}$ valley with spin $\downarrow$ states will be scattered to the $K$ valley with spin $\downarrow$, {via} the assistance of phonon carrying $\mathbf{q}=-K$, together with a chiral phonon emission/absorption of $l_{ph}$ = $-1$ (+1). 
It's worth mentioning that hole scattering with phonons with momentum $\textbf{q}$ can be also achieved at the same time, as shown in Figs.~\ref{fig2}~(a), which will contribute a momentum with $-\mathbf{q}$ for the Ex-Ph process. 

Figure~\ref{fig5}~(d) shows the $\mathbf{q}$-resolved (phonon momentum) Ex-Ph coupling strength with excitons at $K$, which has a relatively high intensity at $\mathbf{q}=K$, confirming the important role that phonon with $\mathbf{q}=K$ play{s} in the photoexcited valley polarization process. 
Thus, via $\sigma^-/\sigma^+$ modulation, photoexcited valley polarization together with a spin-valley locking effect can be obtained in MoSi$_2$N$_4$, which can help to manipulate the diffusion process in photoelectric devices and realize information encoding in exciton transistors.

\section*{Conclusion}

{
    The non-adiabatic effects and the dynamics of the excitons are barely discussed before. But in this work, they are successfully simulated and investigated with our self-developed codes. We also study the long-neglected excitons with non-zero momentum, which are found to be crucial in the exciton dynamics. Time-resolved Ex-Ph dynamics is studied by rt-BTE for the first time, i.e., the evolution of excitons after the photoexcitation, via our self-developed code.
}

We conduct a comprehensive study on the non-adiabatic properties of exciton in MoSi$_2$N$_4$ for the first time. 
Starting from the analysis of the PL process mostly contributed by the Ex-Ph coupling of dark excitons, we unveil their contributions to the bright-exciton-photon coupling behaviors and discover their strong Ex-Ph scattering rates over the whole BZ, with values ranging from 10$^{-2}$ fs$^{-1}$ to 10$^1$ fs$^{-1}$. 
We notice that at $\Lambda$ valley, dark excitons have negative effective masses and some of them have the strongest Ex-Ph coupling effects, which originate from their distinctive intravalley and intervalley scattering processes. 
These findings can help us to have more insights into further exciton-related investigations, especially for the barely studied dark excitons.

The time-resolved Ex-Ph dynamic results show that the whole system needs about 200 fs to get back to the equilibrium and $\Gamma\to K/K^{\prime}$ intervalley scatterings are essential in this process. 
We also propose a new way to modulate the Ex-Ph coupling and valley polarization, together with spin-valley locking effects, by using the photon circular dichroism properties of photoexcitation, based on the conservation rules of spin, energy, momentum and pseudo-angular momentum in the electron-phonon scattering process. 
Both the non-adiabatic exciton dynamics and the spin-valley exciton dynamics at $K/K'$ will inspire scientists to explore more exciton-related physics in MoSi$_2$N$_4$, which will promote the development of their practical applications.

\section*{Acknowledgement}
We acknowledge the support from Tokodai Institute for Element Strategy (TIES) funded by MEXT Elements Strategy Initiative to Form Core Research Center Grants No. JPMXP0112101001, and from JSPS KAKENHI grants JP18H03678 and JP20H04633. 
T. Z. also acknowledges the support by the Japan Society for the Promotion of Science (JSPS), KAKENHI Grant No. JP21K13865.
\section*{Credit author statement}
T.Z. devised the project idea. M.K. performed most of the calculations. S.M. and T.Z. analyzed the results. All the authors prepared the main part of the manuscript and edited the manuscript.
\section*{Declaration of competing interest}
The Authors declare no Competing Financial or Non-Financial Interests.
\section*{Methods}
\subsection{Exciton adiabatic state calculation based on GW-BSE method}

In real materials, their excitonic properties are strongly related to their electronic structures, especially their band gap. 
Therefore, the GW calculation is set to be the first step to study exciton physics, since it can offer a precise band structure of quasiparticles including the complicated electron-electron many-body effects. 
An one-shot $G_0W_0$ calculation is conducted in our work, in which the formula of self-energy correction to the DFT eigenvalues can be written as,

\begin{equation}
E_{n\textbf{k}}^{QP}=\epsilon_{n\textbf{k}} + \bra{\psi_{n\textbf{k}}}\Sigma(E_{n\textbf{k}}^{QP})-V_{xc} \ket{\psi_{n\textbf{k}}},
\end{equation}
where $\epsilon_{n\textbf{k}}$ is the DFT eigenvalue given by the exchange-correlation potential $V_{xc}$ in the local density approximation (LDA). Here, the generalized plasmon pole (GPP) model is used to calculate the self-energy term in our calculation. 

The BSE kernel builds a bridge between bare exciton propagator and bare electron-hole propagator by including the effects of the bare exchange Coulomb repulsion and screened Coulomb attraction. The kernel can be expressed as a two-particle Hamiltonian matrix,

\begin{equation}
H_{vc\textbf{k},v'c'\textbf{k}'} = (\epsilon_{c\textbf{k}}-\epsilon_{v\textbf{k}})\delta_{vv'}\delta_{cc'}\delta_{\textbf{k}\textbf{k}'}+(f_{c\textbf{k}}-f_{v\textbf{k}})\Big{[} K^{x} + K^{c} \Big{]},
\end{equation}
where $\epsilon_{v\textbf{k}}$/$\epsilon_{c\textbf{k}}$ is the eigenvalue of the hole/electron, $f$ is the occupation number and the last two kernel terms $K^{x}$ and $K^{c}$ are the exchange and attraction part, respectively. 
By diagonalizing this BSE matrix, we can get the exciton states with each momentum $\textbf{Q}$ in the adiabatic approximation, which are the stepping stones for our further non-adiabatic investigation.

\subsection{First-order Ex-Ph coupling under Fan-Migdal approximation}

The 1st-order Ex-Ph coupling in materials is mainly contributed by the Fan-Migdal and the Debye-Waller self-energy terms and the dumping effects are only caused by the Fan-Migdal terms.
Here, we discuss the 1st-order Ex-Ph coupling under Fan-Migdal approximation and the first-order exciton-phonon coupling matrix elements can be written as,

\begin{equation}
\begin{split}
\mathcal{G}_{SS'\nu}(\textbf{Q},\textbf{q}) = \sum_{vcc'\textbf{k}} A_{vc\textbf{k}}^{S'\textbf{Q}+\textbf{q}*} A_{vc'\textbf{k}}^{S\textbf{Q}} g_{cc'\nu}(\textbf{k}+\textbf{Q},\textbf{q})\\
-\sum_{vv'c\textbf{k}} A_{vc\textbf{k}}^{S'\textbf{Q}+\textbf{q}*} A_{v'c\textbf{k}+\textbf{q}}^{S\textbf{Q}} g_{vv'\nu}(\textbf{k}+\textbf{q}),
\label{ExPhMatrix}
\end{split}
\end{equation}

where the $A_{vc\textbf{k}}^{S\textbf{Q}}$ is the projection coefficent from exciton state $\ket{\textbf{S}}$ onto electron-hole basis $\ket{vc}$, $g_{mn\lambda}(\textbf{k},\textbf{q})$ is the first-order EPC matrix element, where $m/n$ is the electron band index, $\nu/\nu'$ is the phonon branch index and $\textbf{k}/\textbf{q}$ is the momentum for hole/phonon. The first-order EPI matrix is implemented with the density functional perturbation theory (DFPT), which is treated as a good approximation to the GW perturabtion theory (GWPT). With the first-order exciton-phonon coupling matrix, the dynamical Fan-Migdal (FMd) self-energy can be given as,

\begin{equation}
\begin{split}
\Sigma_{S}^{FMd}(i\omega_n) = \sum_{\nu}\sum_{S'} |\mathcal{G}_{SS'\nu}|^2 \times \Big{[} &\frac{N_B(\omega_\nu)-N_B(\Omega_{S'})}{i\omega_n-\Omega_{S'}+\omega_\nu}\\
&+ \frac{N_B(\omega_\nu+1+N_B(\Omega_{S'}))}{i\omega_n-\Omega_{S'}-\omega_\nu} \Big{]},
\end{split}
\end{equation}

where $N_B$ is the Bose-Einstein distribution and the $i\omega_n$ will be replaced by $i\omega_n = \omega + i\eta$, in which $\eta$ is the given broadening. The static Fan-Migdal (FMs) term $\Sigma_{SS'}^{FMs}$ is neglected here because it only includes the transitions which are virtual and there'll be no contribution to the renormalization of excitons' energy and excitons' lifetime. The Debye-Waller terms' effect on self-energy correction is temporarily not discussed here and will be included in future works. Besides, the non-excitonic terms are neglected here and only the Fan-Migdal term contributes to the excitons' lifetime. Thus the excitons' lifetime led by nonadiatic effect can be simply expressed under SERTA as~\cite{Chen2021},

\begin{equation}
\begin{split}
\frac{1}{\tau_{S\textbf{Q}}} = \frac{2\pi}{\hbar}\frac{1}{N_\textbf{q}}& \sum_{S'\nu\textbf{q}} |\mathcal{G}_{SS'\nu}(\textbf{Q},\textbf{q})|^2\\ \times[(N_{\nu\textbf{q}}&+1+F_{m\textbf{Q}+\textbf{q}})\times\delta(E_{n\textbf{Q}}-E'_{m\textbf{Q}+\textbf{q}}-\hbar\omega_{\nu\textbf{q}})
\\&+(N_{\nu\textbf{q}}-F_{m\textbf{Q}+\textbf{q}})\times\delta(E_{n\textbf{Q}}-E'_{m\textbf{Q}+\textbf{q}}+\hbar\omega_{\nu\textbf{q}})],
\end{split}
\end{equation}

and with the exciton-phonon coupling strength formula, the phonon-assisted PL spectrum can be predicted as\cite{Chen2021},

\begin{equation}
\begin{split}
I(\omega) \propto \sum_{mn\nu}|p_{S_m}|^2 &\int d\textbf{Q} |\mathcal{G}_{nm\nu}(\textbf{Q},-\textbf{Q})|^2 \\&\cdot N_{n\textbf{Q}} \frac{1+N(\hbar\omega_{\nu\textbf{Q}})}{(\hbar\omega-E_m)^2} \cdot \delta(\hbar\omega+\hbar\omega_{\nu\textbf{Q}}-E_{n\textbf{Q}}).
\label{PL}
\end{split}
\end{equation}

\subsection{Computational details}

With the help of PWSCF and PHONON distributions in Quantum ESPRESSO~\cite{giannozzi2009}, the spinful electronic structure is calculated with a 24$\times$24$\times$1 $k$-mesh grid and Optimized Norm-Conserving Vanderbilt (ONCV) pseudo-potential~\cite{Martin2015} by DFT, while the phonon spectra is calculated with a 6$\times$6$\times$1 $k$-mesh and 2D Born-charge correction at $\Gamma$ point by DFPT. 
One-shot $G_0W_0$ calculation is implemented with the same $k$-mesh as DFT calculation and the GPP algorithm for QP calculation of 8 bands on both sides of the Fermi level by YAMBO~\cite{Andrea2009,Sangalli2019}. The BSE calculation is also performed with YAMBO code. 

In the Ex-Ph coupling part, the numerical calculation is based on our self-developed code. MPI is used for parallel computation with high efficiency. We use the same $k$-meshes (24$\times$24$\times$1) for the calculations of electron, phonon and exciton. 
Eigenstates for electrons/holes on the closest two bands of each side of the forbidden gap are chosen as the basis of exciton envelop wave function according to the materials' electronic structure pattern. The Ex-Ph coupling is only calculated with the lowest 15 exciton bands due to their low occupation numbers and relatively strong coupling effects. And the rt-BTE method is based on the formula given in Ref.~\cite{Bernardi2016}.

{
    Here, we conduct the calculation on a, strictly speaking, bulk but optimized stable crystal structure of $MoSi_2N_4$. Studies have shown that this material is quasi-2D since there is little difference between the band structure of the bulk, monolayer and multilayer structures~\cite{Wu2021}. Moreover, different structures turn out to have almost the same valley properties, which means that the exciton-related properties with different geometries of this material will not change significantly in our study.
}
{
    We also note that the phonon spectra in our paper are slightly different from the results given in Ref~\cite{Yu_2021}, which is mainly caused by the spin-orbit coupling effect (SOC), the different codes and the different pseudopotentials. The acoustic rule can not be fully ensured to exact zero leading to the poor interpolation around the $\Gamma$ point. In Ref~\cite{Yu_2021}, the calculation is performed with PAW pseudopotential in VASP (Vienna Ab-initio Simulation Packages) while we use the ONCV pseudopotentials in Quantum ESPRESSO.
}

%%%END OF MAIN TEXT%%%

%  For footnotes in the main text of the article please number the footnotes to avoid duplicate symbols. e.g.  \footnote[num]{your text} the corresponding author \ast counts as footnote 1, ESI as footnote 2, e.g. if there is no ESI, please start at [num]=[2], if ESI is cited in the title please start at [num]=[3] etc. Please also cite the ESI within the main body of the text using \dag.

% The \balance command can be used to balance the columns on the final page if desired. It should be placed anywhere within the first column of the last page.

% \balance

% If notes are included in your references you can change the title from 'References' to 'Notes and references' using the following command:
% \renewcommand\refname{Notes and references}

%%%REFERENCES%%%
\scriptsize{
\bibliography{ref} %You need to replace "rsc" on this line with the name of your .bib file

\providecommand*{\mcitethebibliography}{\thebibliography}
\csname @ifundefined\endcsname{endmcitethebibliography}
{\let\endmcitethebibliography\endthebibliography}{}
\begin{mcitethebibliography}{67}
\providecommand*{\natexlab}[1]{#1}
\providecommand*{\mciteSetBstSublistMode}[1]{}
\providecommand*{\mciteSetBstMaxWidthForm}[2]{}
\providecommand*{\mciteBstWouldAddEndPuncttrue}
  {\def\EndOfBibitem{\unskip.}}
\providecommand*{\mciteBstWouldAddEndPunctfalse}
  {\let\EndOfBibitem\relax}
\providecommand*{\mciteSetBstMidEndSepPunct}[3]{}
\providecommand*{\mciteSetBstSublistLabelBeginEnd}[3]{}
\providecommand*{\EndOfBibitem}{}
\mciteSetBstSublistMode{f}
\mciteSetBstMaxWidthForm{subitem}
{(\emph{\alph{mcitesubitemcount}})}
\mciteSetBstSublistLabelBeginEnd{\mcitemaxwidthsubitemform\space}
{\relax}{\relax}

\bibitem[Wang \emph{et~al.}(2020)Wang, Nie, Li, Zuo, Fauqué, Zhu, and
  Behnia]{Jinhua2020}
J.~Wang, P.~Nie, X.~Li, H.~Zuo, B.~Fauqué, Z.~Zhu and K.~Behnia,
  \emph{Proceedings of the National Academy of Sciences}, 2020, \textbf{117},
  30215--30219\relax
\mciteBstWouldAddEndPuncttrue
\mciteSetBstMidEndSepPunct{\mcitedefaultmidpunct}
{\mcitedefaultendpunct}{\mcitedefaultseppunct}\relax
\EndOfBibitem
\bibitem[Guerci \emph{et~al.}(2019)Guerci, Capone, and Fabrizio]{Guerci2019}
D.~Guerci, M.~Capone and M.~Fabrizio, \emph{Phys. Rev. Materials}, 2019,
  \textbf{3}, 054605\relax
\mciteBstWouldAddEndPuncttrue
\mciteSetBstMidEndSepPunct{\mcitedefaultmidpunct}
{\mcitedefaultendpunct}{\mcitedefaultseppunct}\relax
\EndOfBibitem
\bibitem[B\"oning \emph{et~al.}(2011)B\"oning, Filinov, and Bonitz]{Boning2011}
J.~B\"oning, A.~Filinov and M.~Bonitz, \emph{Phys. Rev. B}, 2011, \textbf{84},
  075130\relax
\mciteBstWouldAddEndPuncttrue
\mciteSetBstMidEndSepPunct{\mcitedefaultmidpunct}
{\mcitedefaultendpunct}{\mcitedefaultseppunct}\relax
\EndOfBibitem
\bibitem[Wu \emph{et~al.}(2022)Wu, Jiang, Chen, Liu, Liu, and Xie]{Wu2022}
Y.~Wu, H.~Jiang, H.~Chen, H.~Liu, J.~Liu and X.~C. Xie, \emph{Phys. Rev.
  Lett.}, 2022, \textbf{128}, 106804\relax
\mciteBstWouldAddEndPuncttrue
\mciteSetBstMidEndSepPunct{\mcitedefaultmidpunct}
{\mcitedefaultendpunct}{\mcitedefaultseppunct}\relax
\EndOfBibitem
\bibitem[Menke \emph{et~al.}(2013)Menke, Luhman, and Holmes]{Menke2013}
S.~M. Menke, W.~A. Luhman and R.~J. Holmes, \emph{Nature Materials}, 2013,
  \textbf{12}, 152--157\relax
\mciteBstWouldAddEndPuncttrue
\mciteSetBstMidEndSepPunct{\mcitedefaultmidpunct}
{\mcitedefaultendpunct}{\mcitedefaultseppunct}\relax
\EndOfBibitem
\bibitem[Zhang \emph{et~al.}(2019)Zhang, Dement, Ferry, and Holmes]{Zhang2019}
T.~Zhang, D.~B. Dement, V.~E. Ferry and R.~J. Holmes, \emph{Nature
  Communications}, 2019, \textbf{10}, 1156\relax
\mciteBstWouldAddEndPuncttrue
\mciteSetBstMidEndSepPunct{\mcitedefaultmidpunct}
{\mcitedefaultendpunct}{\mcitedefaultseppunct}\relax
\EndOfBibitem
\bibitem[Ehrler \emph{et~al.}(2012)Ehrler, Walker, B{\"o}hm, Wilson, Vaynzof,
  Friend, and Greenham]{Ehrler2012}
B.~Ehrler, B.~J. Walker, M.~L. B{\"o}hm, M.~W. Wilson, Y.~Vaynzof, R.~H. Friend
  and N.~C. Greenham, \emph{Nature Communications}, 2012, \textbf{3},
  1019\relax
\mciteBstWouldAddEndPuncttrue
\mciteSetBstMidEndSepPunct{\mcitedefaultmidpunct}
{\mcitedefaultendpunct}{\mcitedefaultseppunct}\relax
\EndOfBibitem
\bibitem[Hedley \emph{et~al.}(2013)Hedley, Ward, Alekseev, Howells, Martins,
  Serrano, Cooke, Ruseckas, and Samuel]{Hedley2013}
G.~J. Hedley, A.~J. Ward, A.~Alekseev, C.~T. Howells, E.~R. Martins, L.~A.
  Serrano, G.~Cooke, A.~Ruseckas and I.~D.~W. Samuel, \emph{Nature
  Communications}, 2013, \textbf{4}, 2867\relax
\mciteBstWouldAddEndPuncttrue
\mciteSetBstMidEndSepPunct{\mcitedefaultmidpunct}
{\mcitedefaultendpunct}{\mcitedefaultseppunct}\relax
\EndOfBibitem
\bibitem[Chen \emph{et~al.}(2021)Chen, Atallah, Lin, Wang, Lee, Xu, Huang,
  Duan, Ping, Huang, Caram, and Duan]{Chen2021}
P.~Chen, T.~L. Atallah, Z.~Lin, P.~Wang, S.-J. Lee, J.~Xu, Z.~Huang, X.~Duan,
  Y.~Ping, Y.~Huang, J.~R. Caram and X.~Duan, \emph{Nature}, 2021,
  \textbf{599}, 404--410\relax
\mciteBstWouldAddEndPuncttrue
\mciteSetBstMidEndSepPunct{\mcitedefaultmidpunct}
{\mcitedefaultendpunct}{\mcitedefaultseppunct}\relax
\EndOfBibitem
\bibitem[Mesta \emph{et~al.}(2013)Mesta, Carvelli, de~Vries, van Eersel,
  van~der Holst, Schober, Furno, L{\"u}ssem, Leo, Loebl, Coehoorn, and
  Bobbert]{Mesta2013}
M.~Mesta, M.~Carvelli, R.~J. de~Vries, H.~van Eersel, J.~J.~M. van~der Holst,
  M.~Schober, M.~Furno, B.~L{\"u}ssem, K.~Leo, P.~Loebl, R.~Coehoorn and P.~A.
  Bobbert, \emph{Nature Materials}, 2013, \textbf{12}, 652--658\relax
\mciteBstWouldAddEndPuncttrue
\mciteSetBstMidEndSepPunct{\mcitedefaultmidpunct}
{\mcitedefaultendpunct}{\mcitedefaultseppunct}\relax
\EndOfBibitem
\bibitem[Hasan \emph{et~al.}(2022)Hasan, Saggar, Shukla, Bencheikh, Sobus,
  McGregor, Adachi, Lo, and Namdas]{Hasan2022}
M.~Hasan, S.~Saggar, A.~Shukla, F.~Bencheikh, J.~Sobus, S.~K.~M. McGregor,
  C.~Adachi, S.-C. Lo and E.~B. Namdas, \emph{Nature Communications}, 2022,
  \textbf{13}, 254\relax
\mciteBstWouldAddEndPuncttrue
\mciteSetBstMidEndSepPunct{\mcitedefaultmidpunct}
{\mcitedefaultendpunct}{\mcitedefaultseppunct}\relax
\EndOfBibitem
\bibitem[Hofmann \emph{et~al.}(2012)Hofmann, Rosenow, Gather, L{\"u}ssem, and
  Leo]{Hofmann2012}
S.~Hofmann, T.~C. Rosenow, M.~C. Gather, B.~L{\"u}ssem and K.~Leo,
  \emph{Physical Review B}, 2012, \textbf{85}, 245209\relax
\mciteBstWouldAddEndPuncttrue
\mciteSetBstMidEndSepPunct{\mcitedefaultmidpunct}
{\mcitedefaultendpunct}{\mcitedefaultseppunct}\relax
\EndOfBibitem
\bibitem[Gramlich \emph{et~al.}(2021)Gramlich, Lampe, Drewniok, and
  Urban]{Gramlich2021}
M.~Gramlich, C.~Lampe, J.~Drewniok and A.~S. Urban, \emph{The Journal of
  Physical Chemistry Letters}, 2021, \textbf{12}, 11371--11377\relax
\mciteBstWouldAddEndPuncttrue
\mciteSetBstMidEndSepPunct{\mcitedefaultmidpunct}
{\mcitedefaultendpunct}{\mcitedefaultseppunct}\relax
\EndOfBibitem
\bibitem[Rodina and Efros(2016)]{Rodina2016}
A.~V. Rodina and A.~L. Efros, \emph{Phys. Rev. B}, 2016, \textbf{93},
  155427\relax
\mciteBstWouldAddEndPuncttrue
\mciteSetBstMidEndSepPunct{\mcitedefaultmidpunct}
{\mcitedefaultendpunct}{\mcitedefaultseppunct}\relax
\EndOfBibitem
\bibitem[Zhang \emph{et~al.}(2022)Zhang, Sung, Toolan, Han, Pandya, Weir, Xiao,
  Dowland, Liu, Ryan, Jones, Huang, and Rao]{Zhang2022}
Z.~Zhang, J.~Sung, D.~T.~W. Toolan, S.~Han, R.~Pandya, M.~P. Weir, J.~Xiao,
  S.~Dowland, M.~Liu, A.~J. Ryan, R.~A.~L. Jones, S.~Huang and A.~Rao,
  \emph{Nature Materials}, 2022\relax
\mciteBstWouldAddEndPuncttrue
\mciteSetBstMidEndSepPunct{\mcitedefaultmidpunct}
{\mcitedefaultendpunct}{\mcitedefaultseppunct}\relax
\EndOfBibitem
\bibitem[Gupta \emph{et~al.}(2021)Gupta, Bitton, Neuman, Esteban, Chuntonov,
  Aizpurua, and Haran]{Gupta2021}
S.~N. Gupta, O.~Bitton, T.~Neuman, R.~Esteban, L.~Chuntonov, J.~Aizpurua and
  G.~Haran, \emph{Nature Communications}, 2021, \textbf{12}, 1310\relax
\mciteBstWouldAddEndPuncttrue
\mciteSetBstMidEndSepPunct{\mcitedefaultmidpunct}
{\mcitedefaultendpunct}{\mcitedefaultseppunct}\relax
\EndOfBibitem
\bibitem[Unuchek \emph{et~al.}(2018)Unuchek, Ciarrocchi, Avsar, Watanabe,
  Taniguchi, and Kis]{Unuchek2018}
D.~Unuchek, A.~Ciarrocchi, A.~Avsar, K.~Watanabe, T.~Taniguchi and A.~Kis,
  \emph{Nature}, 2018, \textbf{560}, 340--344\relax
\mciteBstWouldAddEndPuncttrue
\mciteSetBstMidEndSepPunct{\mcitedefaultmidpunct}
{\mcitedefaultendpunct}{\mcitedefaultseppunct}\relax
\EndOfBibitem
\bibitem[Zhang and Murakami(2022)]{zhang2022chiral}
T.~Zhang and S.~Murakami, \emph{Physical {R}eview {B}}, 2022, \textbf{105},
  235204\relax
\mciteBstWouldAddEndPuncttrue
\mciteSetBstMidEndSepPunct{\mcitedefaultmidpunct}
{\mcitedefaultendpunct}{\mcitedefaultseppunct}\relax
\EndOfBibitem
\bibitem[Ciarrocchi \emph{et~al.}(2019)Ciarrocchi, Unuchek, Avsar, Watanabe,
  Taniguchi, and Kis]{Ciarrocchi2019}
A.~Ciarrocchi, D.~Unuchek, A.~Avsar, K.~Watanabe, T.~Taniguchi and A.~Kis,
  \emph{Nature Photonics}, 2019, \textbf{13}, 131--136\relax
\mciteBstWouldAddEndPuncttrue
\mciteSetBstMidEndSepPunct{\mcitedefaultmidpunct}
{\mcitedefaultendpunct}{\mcitedefaultseppunct}\relax
\EndOfBibitem
\bibitem[Marini(2008)]{Marini2008}
A.~Marini, \emph{Phys. Rev. Lett.}, 2008, \textbf{101}, 106405\relax
\mciteBstWouldAddEndPuncttrue
\mciteSetBstMidEndSepPunct{\mcitedefaultmidpunct}
{\mcitedefaultendpunct}{\mcitedefaultseppunct}\relax
\EndOfBibitem
\bibitem[Cannuccia \emph{et~al.}(2019)Cannuccia, Monserrat, and
  Attaccalite]{Cannuccia2019}
E.~Cannuccia, B.~Monserrat and C.~Attaccalite, \emph{Phys. Rev. B}, 2019,
  \textbf{99}, 081109\relax
\mciteBstWouldAddEndPuncttrue
\mciteSetBstMidEndSepPunct{\mcitedefaultmidpunct}
{\mcitedefaultendpunct}{\mcitedefaultseppunct}\relax
\EndOfBibitem
\bibitem[Paleari \emph{et~al.}(2019)Paleari, P.~C.~Miranda, Molina-S\'anchez,
  and Wirtz]{Paleari2019}
F.~Paleari, H.~P.~C.~Miranda, A.~Molina-S\'anchez and L.~Wirtz, \emph{Phys.
  Rev. Lett.}, 2019, \textbf{122}, 187401\relax
\mciteBstWouldAddEndPuncttrue
\mciteSetBstMidEndSepPunct{\mcitedefaultmidpunct}
{\mcitedefaultendpunct}{\mcitedefaultseppunct}\relax
\EndOfBibitem
\bibitem[Chen \emph{et~al.}(2020)Chen, Sangalli, and Bernardi]{Chen2020}
H.-Y. Chen, D.~Sangalli and M.~Bernardi, \emph{Phys. Rev. Lett.}, 2020,
  \textbf{125}, 107401\relax
\mciteBstWouldAddEndPuncttrue
\mciteSetBstMidEndSepPunct{\mcitedefaultmidpunct}
{\mcitedefaultendpunct}{\mcitedefaultseppunct}\relax
\EndOfBibitem
\bibitem[Zhang \emph{et~al.}(2021)Zhang, Xie, Wang, Cao, and Li]{Zhang2021}
X.-W. Zhang, K.~Xie, E.-G. Wang, T.~Cao and X.-Z. Li, \emph{Phonon-mediated
  exciton relaxation in two-dimensional semiconductors: selection rules and
  relaxation pathways}, 2021, \url{https://arxiv.org/abs/2110.08873}\relax
\mciteBstWouldAddEndPuncttrue
\mciteSetBstMidEndSepPunct{\mcitedefaultmidpunct}
{\mcitedefaultendpunct}{\mcitedefaultseppunct}\relax
\EndOfBibitem
\bibitem[Jiang \emph{et~al.}(2021)Jiang, Zheng, Lan, Saidi, Ren, and
  Zhao]{Xiang2021}
X.~Jiang, Q.~Zheng, Z.~Lan, W.~A. Saidi, X.~Ren and J.~Zhao, \emph{Science
  Advances}, 2021, \textbf{7}, eabf3759\relax
\mciteBstWouldAddEndPuncttrue
\mciteSetBstMidEndSepPunct{\mcitedefaultmidpunct}
{\mcitedefaultendpunct}{\mcitedefaultseppunct}\relax
\EndOfBibitem
\bibitem[Antonius and Louie(2022)]{Antonius2022}
G.~Antonius and S.~G. Louie, \emph{Phys. Rev. B}, 2022, \textbf{105},
  085111\relax
\mciteBstWouldAddEndPuncttrue
\mciteSetBstMidEndSepPunct{\mcitedefaultmidpunct}
{\mcitedefaultendpunct}{\mcitedefaultseppunct}\relax
\EndOfBibitem
\bibitem[Linardy \emph{et~al.}(2020)Linardy, Yadav, Vella, Verzhbitskiy,
  Watanabe, Taniguchi, Pauly, Trushin, and Eda]{Linardy2020}
E.~Linardy, D.~Yadav, D.~Vella, I.~A. Verzhbitskiy, K.~Watanabe, T.~Taniguchi,
  F.~Pauly, M.~Trushin and G.~Eda, \emph{Nano Letters}, 2020, \textbf{20},
  1647--1653\relax
\mciteBstWouldAddEndPuncttrue
\mciteSetBstMidEndSepPunct{\mcitedefaultmidpunct}
{\mcitedefaultendpunct}{\mcitedefaultseppunct}\relax
\EndOfBibitem
\bibitem[Erkensten \emph{et~al.}(2021)Erkensten, Brem, and
  Malic]{Erkensten2021}
D.~Erkensten, S.~Brem and E.~Malic, \emph{Phys. Rev. B}, 2021, \textbf{103},
  045426\relax
\mciteBstWouldAddEndPuncttrue
\mciteSetBstMidEndSepPunct{\mcitedefaultmidpunct}
{\mcitedefaultendpunct}{\mcitedefaultseppunct}\relax
\EndOfBibitem
\bibitem[Efimkin \emph{et~al.}(2021)Efimkin, Laird, Levinsen, Parish, and
  MacDonald]{Efimkin2021}
D.~K. Efimkin, E.~K. Laird, J.~Levinsen, M.~M. Parish and A.~H. MacDonald,
  \emph{Phys. Rev. B}, 2021, \textbf{103}, 075417\relax
\mciteBstWouldAddEndPuncttrue
\mciteSetBstMidEndSepPunct{\mcitedefaultmidpunct}
{\mcitedefaultendpunct}{\mcitedefaultseppunct}\relax
\EndOfBibitem
\bibitem[Steinhoff \emph{et~al.}(2017)Steinhoff, Florian, R{\"o}sner,
  Sch{\"o}nhoff, Wehling, and Jahnke]{Steinhoff2017}
A.~Steinhoff, M.~Florian, M.~R{\"o}sner, G.~Sch{\"o}nhoff, T.~O. Wehling and
  F.~Jahnke, \emph{Nature Communications}, 2017, \textbf{8}, 1166\relax
\mciteBstWouldAddEndPuncttrue
\mciteSetBstMidEndSepPunct{\mcitedefaultmidpunct}
{\mcitedefaultendpunct}{\mcitedefaultseppunct}\relax
\EndOfBibitem
\bibitem[Ugeda \emph{et~al.}(2014)Ugeda, Bradley, Shi, da~Jornada, Zhang, Qiu,
  Ruan, Mo, Hussain, Shen, Wang, Louie, and Crommie]{Ugeda2014}
M.~M. Ugeda, A.~J. Bradley, S.-F. Shi, F.~H. da~Jornada, Y.~Zhang, D.~Y. Qiu,
  W.~Ruan, S.-K. Mo, Z.~Hussain, Z.-X. Shen, F.~Wang, S.~G. Louie and M.~F.
  Crommie, \emph{Nature Materials}, 2014, \textbf{13}, 1091--1095\relax
\mciteBstWouldAddEndPuncttrue
\mciteSetBstMidEndSepPunct{\mcitedefaultmidpunct}
{\mcitedefaultendpunct}{\mcitedefaultseppunct}\relax
\EndOfBibitem
\bibitem[Chernikov \emph{et~al.}(2014)Chernikov, Berkelbach, Hill, Rigosi, Li,
  Aslan, Reichman, Hybertsen, and Heinz]{Chernikov2014}
A.~Chernikov, T.~C. Berkelbach, H.~M. Hill, A.~Rigosi, Y.~Li, O.~B. Aslan,
  D.~R. Reichman, M.~S. Hybertsen and T.~F. Heinz, \emph{Phys. Rev. Lett.},
  2014, \textbf{113}, 076802\relax
\mciteBstWouldAddEndPuncttrue
\mciteSetBstMidEndSepPunct{\mcitedefaultmidpunct}
{\mcitedefaultendpunct}{\mcitedefaultseppunct}\relax
\EndOfBibitem
\bibitem[Hong \emph{et~al.}(2020)Hong, Liu, Wang, Zhou, Ma, Xu, Feng, Chen,
  Chen, Sun, Chen, Cheng, and Ren]{YiLun2020}
Y.-L. Hong, Z.~Liu, L.~Wang, T.~Zhou, W.~Ma, C.~Xu, S.~Feng, L.~Chen, M.-L.
  Chen, D.-M. Sun, X.-Q. Chen, H.-M. Cheng and W.~Ren, \emph{Science}, 2020,
  \textbf{369}, 670--674\relax
\mciteBstWouldAddEndPuncttrue
\mciteSetBstMidEndSepPunct{\mcitedefaultmidpunct}
{\mcitedefaultendpunct}{\mcitedefaultseppunct}\relax
\EndOfBibitem
\bibitem[Wu \emph{et~al.}(2021)Wu, Tang, Xia, Gao, Jia, Zhang, Zhu, Zhang, and
  Zhang]{Wu2021}
Y.~Wu, Z.~Tang, W.~Xia, W.~Gao, F.~Jia, Y.~Zhang, W.~Zhu, W.~Zhang and
  P.~Zhang, \emph{MoSi2N4: An emerging 2D electronic material with protected
  band edge states and dielectric tunable quasiparticle and optical
  properties}, 2021, \url{https://arxiv.org/abs/2107.10126}\relax
\mciteBstWouldAddEndPuncttrue
\mciteSetBstMidEndSepPunct{\mcitedefaultmidpunct}
{\mcitedefaultendpunct}{\mcitedefaultseppunct}\relax
\EndOfBibitem
\bibitem[Hong \emph{et~al.}(2020)Hong, Liu, Wang, Zhou, Ma, Xu, Feng, Chen,
  Chen, Sun, Chen, Cheng, and Ren]{Hong2020}
Y.-L. Hong, Z.~Liu, L.~Wang, T.~Zhou, W.~Ma, C.~Xu, S.~Feng, L.~Chen, M.-L.
  Chen, D.-M. Sun, X.-Q. Chen, H.-M. Cheng and W.~Ren, \emph{Science}, 2020,
  \textbf{369}, 670--674\relax
\mciteBstWouldAddEndPuncttrue
\mciteSetBstMidEndSepPunct{\mcitedefaultmidpunct}
{\mcitedefaultendpunct}{\mcitedefaultseppunct}\relax
\EndOfBibitem
\bibitem[Novoselov(2020)]{Novoselov2020}
K.~S. Novoselov, \emph{National Science Review}, 2020, \textbf{7},
  1842--1844\relax
\mciteBstWouldAddEndPuncttrue
\mciteSetBstMidEndSepPunct{\mcitedefaultmidpunct}
{\mcitedefaultendpunct}{\mcitedefaultseppunct}\relax
\EndOfBibitem
\bibitem[Li \emph{et~al.}(2020)Li, Wu, Feng, Guan, Feng, Yao, and
  Yang]{li2020valley}
S.~Li, W.~Wu, X.~Feng, S.~Guan, W.~Feng, Y.~Yao and S.~A. Yang, \emph{Physical
  {R}eview {B}}, 2020, \textbf{102}, 235435\relax
\mciteBstWouldAddEndPuncttrue
\mciteSetBstMidEndSepPunct{\mcitedefaultmidpunct}
{\mcitedefaultendpunct}{\mcitedefaultseppunct}\relax
\EndOfBibitem
\bibitem[Chen \emph{et~al.}(2021)Chen, Chen, and Zhang]{chen2021first}
R.~Chen, D.~Chen and W.~Zhang, \emph{Results in Physics}, 2021, \textbf{30},
  104864\relax
\mciteBstWouldAddEndPuncttrue
\mciteSetBstMidEndSepPunct{\mcitedefaultmidpunct}
{\mcitedefaultendpunct}{\mcitedefaultseppunct}\relax
\EndOfBibitem
\bibitem[Yao \emph{et~al.}(2021)Yao, Zhang, Wang, Li, Yu, Xu, Wang, and
  Wei]{Yao2021}
H.~Yao, C.~Zhang, Q.~Wang, J.~Li, Y.~Yu, F.~Xu, B.~Wang and Y.~Wei,
  \emph{Nanomaterials 2021, Vol. 11, Page 559}, 2021, \textbf{11}, 559\relax
\mciteBstWouldAddEndPuncttrue
\mciteSetBstMidEndSepPunct{\mcitedefaultmidpunct}
{\mcitedefaultendpunct}{\mcitedefaultseppunct}\relax
\EndOfBibitem
\bibitem[Yang \emph{et~al.}(2021)Yang, Zhao, Shi-Qi, Liu, Wang, Chen, Gao, and
  Zhao]{yang2021accurate}
J.-S. Yang, L.~Zhao, L.~Shi-Qi, H.~Liu, L.~Wang, M.~Chen, J.~Gao and J.~Zhao,
  \emph{Nanoscale}, 2021, \textbf{13}, 5479--5488\relax
\mciteBstWouldAddEndPuncttrue
\mciteSetBstMidEndSepPunct{\mcitedefaultmidpunct}
{\mcitedefaultendpunct}{\mcitedefaultseppunct}\relax
\EndOfBibitem
\bibitem[Wu \emph{et~al.}(2021)Wu, Cao, Ang, and Ang]{wu2021semiconductor}
Q.~Wu, L.~Cao, Y.~S. Ang and L.~K. Ang, \emph{Applied Physics Letters}, 2021,
  \textbf{118}, 113102\relax
\mciteBstWouldAddEndPuncttrue
\mciteSetBstMidEndSepPunct{\mcitedefaultmidpunct}
{\mcitedefaultendpunct}{\mcitedefaultseppunct}\relax
\EndOfBibitem
\bibitem[Zhong \emph{et~al.}(2021)Zhong, Xiong, Lv, Yu, and
  Yuan]{zhong2021strain}
H.~Zhong, W.~Xiong, P.~Lv, J.~Yu and S.~Yuan, \emph{Phys. Rev. B}, 2021,
  \textbf{103}, 085124\relax
\mciteBstWouldAddEndPuncttrue
\mciteSetBstMidEndSepPunct{\mcitedefaultmidpunct}
{\mcitedefaultendpunct}{\mcitedefaultseppunct}\relax
\EndOfBibitem
\bibitem[Bafekry \emph{et~al.}(2021)Bafekry, Faraji, Hoat, Shahrokhi,
  Fadlallah, Shojaei, Feghhi, Ghergherehchi, and Gogova]{Bafekry_2021}
A.~Bafekry, M.~Faraji, D.~M. Hoat, M.~Shahrokhi, M.~M. Fadlallah, F.~Shojaei,
  S.~A.~H. Feghhi, M.~Ghergherehchi and D.~Gogova, 2021, \textbf{54},
  155303\relax
\mciteBstWouldAddEndPuncttrue
\mciteSetBstMidEndSepPunct{\mcitedefaultmidpunct}
{\mcitedefaultendpunct}{\mcitedefaultseppunct}\relax
\EndOfBibitem
\bibitem[Kang and Lin(2021)]{kang2021second}
L.~Kang and Z.~Lin, \emph{Phys. Rev. B}, 2021, \textbf{103}, 195404\relax
\mciteBstWouldAddEndPuncttrue
\mciteSetBstMidEndSepPunct{\mcitedefaultmidpunct}
{\mcitedefaultendpunct}{\mcitedefaultseppunct}\relax
\EndOfBibitem
\bibitem[Cui \emph{et~al.}(2021)Cui, Luo, Yu, and Xu]{cui2021tuning}
Z.~Cui, Y.~Luo, J.~Yu and Y.~Xu, \emph{Physica E: Low-dimensional Systems and
  Nanostructures}, 2021, \textbf{134}, 114873\relax
\mciteBstWouldAddEndPuncttrue
\mciteSetBstMidEndSepPunct{\mcitedefaultmidpunct}
{\mcitedefaultendpunct}{\mcitedefaultseppunct}\relax
\EndOfBibitem
\bibitem[Guo \emph{et~al.}(2020)Guo, Zhu, Mu, and Ren]{guo2020intrinsic}
S.-D. Guo, Y.-T. Zhu, W.-Q. Mu and W.-C. Ren, \emph{EPL (Europhysics Letters)},
  2020, \textbf{132}, 57002\relax
\mciteBstWouldAddEndPuncttrue
\mciteSetBstMidEndSepPunct{\mcitedefaultmidpunct}
{\mcitedefaultendpunct}{\mcitedefaultseppunct}\relax
\EndOfBibitem
\bibitem[Mortazavi \emph{et~al.}(2021)Mortazavi, Javvaji, Shojaei, Rabczuk,
  Shapeev, and Zhuang]{mortazavi2021exceptional}
B.~Mortazavi, B.~Javvaji, F.~Shojaei, T.~Rabczuk, A.~V. Shapeev and X.~Zhuang,
  \emph{Nano Energy}, 2021, \textbf{82}, 105716\relax
\mciteBstWouldAddEndPuncttrue
\mciteSetBstMidEndSepPunct{\mcitedefaultmidpunct}
{\mcitedefaultendpunct}{\mcitedefaultseppunct}\relax
\EndOfBibitem
\bibitem[Guo \emph{et~al.}(2021)Guo, Zhu, Mu, Wang, and Chen]{GUO2021110223}
S.-D. Guo, Y.-T. Zhu, W.-Q. Mu, L.~Wang and X.-Q. Chen, \emph{Computational
  Materials Science}, 2021, \textbf{188}, 110223\relax
\mciteBstWouldAddEndPuncttrue
\mciteSetBstMidEndSepPunct{\mcitedefaultmidpunct}
{\mcitedefaultendpunct}{\mcitedefaultseppunct}\relax
\EndOfBibitem
\bibitem[Li \emph{et~al.}(2021)Li, Zhou, Wan, and Zhou]{LI2021114753}
Q.~Li, W.~Zhou, X.~Wan and J.~Zhou, \emph{Physica E: Low-dimensional Systems
  and Nanostructures}, 2021, \textbf{131}, 114753\relax
\mciteBstWouldAddEndPuncttrue
\mciteSetBstMidEndSepPunct{\mcitedefaultmidpunct}
{\mcitedefaultendpunct}{\mcitedefaultseppunct}\relax
\EndOfBibitem
\bibitem[Wang \emph{et~al.}(2021)Wang, Cao, Liang, Wu, Wang, Lee, Ong, Yang,
  Ang, Yang, and Ang]{wang2021article}
Q.~Wang, L.~Cao, S.-J. Liang, W.~Wu, G.~Wang, C.~H. Lee, W.~L. Ong, H.~Y. Yang,
  L.~K. Ang, S.~A. Yang and Y.~S. Ang, \emph{npj 2D Materials and
  Applications}, 2021, \textbf{5}, 1--9\relax
\mciteBstWouldAddEndPuncttrue
\mciteSetBstMidEndSepPunct{\mcitedefaultmidpunct}
{\mcitedefaultendpunct}{\mcitedefaultseppunct}\relax
\EndOfBibitem
\bibitem[Cao \emph{et~al.}(2021)Cao, Zhou, Wang, Ang, and Ang]{cao2021two}
L.~Cao, G.~Zhou, Q.~Wang, L.~Ang and Y.~S. Ang, \emph{Applied Physics Letters},
  2021, \textbf{118}, 013106\relax
\mciteBstWouldAddEndPuncttrue
\mciteSetBstMidEndSepPunct{\mcitedefaultmidpunct}
{\mcitedefaultendpunct}{\mcitedefaultseppunct}\relax
\EndOfBibitem
\bibitem[Wang \emph{et~al.}(2021)Wang, Shi, Liu, Zhang, Hong, Li, Gao, Chen,
  Ren, Cheng,\emph{et~al.}]{wang2021intercalated}
L.~Wang, Y.~Shi, M.~Liu, A.~Zhang, Y.-L. Hong, R.~Li, Q.~Gao, M.~Chen, W.~Ren,
  H.-M. Cheng \emph{et~al.}, \emph{Nature communications}, 2021, \textbf{12},
  1--10\relax
\mciteBstWouldAddEndPuncttrue
\mciteSetBstMidEndSepPunct{\mcitedefaultmidpunct}
{\mcitedefaultendpunct}{\mcitedefaultseppunct}\relax
\EndOfBibitem
\bibitem[Yu \emph{et~al.}(2021)Yu, Zhou, Wan, and Li]{Yu2021}
J.~Yu, J.~Zhou, X.~Wan and Q.~Li, \emph{New J. Phys}, 2021, \textbf{23},
  33005\relax
\mciteBstWouldAddEndPuncttrue
\mciteSetBstMidEndSepPunct{\mcitedefaultmidpunct}
{\mcitedefaultendpunct}{\mcitedefaultseppunct}\relax
\EndOfBibitem
\bibitem[Nandan \emph{et~al.}(2021)Nandan, Ghosh, Agarwal, Bhowmick, and
  Chauhan]{nandan2021two}
K.~Nandan, B.~Ghosh, A.~Agarwal, S.~Bhowmick and Y.~S. Chauhan, \emph{IEEE
  Transactions on Electron Devices}, 2021, \textbf{69}, 406--413\relax
\mciteBstWouldAddEndPuncttrue
\mciteSetBstMidEndSepPunct{\mcitedefaultmidpunct}
{\mcitedefaultendpunct}{\mcitedefaultseppunct}\relax
\EndOfBibitem
\bibitem[Sohier \emph{et~al.}(2017)Sohier, Gibertini, Calandra, Mauri, and
  Marzari]{Sohier2017}
T.~Sohier, M.~Gibertini, M.~Calandra, F.~Mauri and N.~Marzari, \emph{Nano
  Letters}, 2017, \textbf{17}, 3758--3763\relax
\mciteBstWouldAddEndPuncttrue
\mciteSetBstMidEndSepPunct{\mcitedefaultmidpunct}
{\mcitedefaultendpunct}{\mcitedefaultseppunct}\relax
\EndOfBibitem
\bibitem[Huang \emph{et~al.}()Huang, Liang, Guo, Lu, Wang, Yu, and
  Zhang]{Huang2022}
D.~Huang, F.~Liang, R.~Guo, D.~Lu, J.~Wang, H.~Yu and H.~Zhang, \emph{Advanced
  Optical Materials}, \textbf{n/a}, 2102612\relax
\mciteBstWouldAddEndPuncttrue
\mciteSetBstMidEndSepPunct{\mcitedefaultmidpunct}
{\mcitedefaultendpunct}{\mcitedefaultseppunct}\relax
\EndOfBibitem
\bibitem[Liang \emph{et~al.}(2022)Liang, Xu, Lu, and Cai]{Liang2022}
D.~Liang, S.~Xu, P.~Lu and Y.~Cai, \emph{Phys. Rev. B}, 2022, \textbf{105},
  195302\relax
\mciteBstWouldAddEndPuncttrue
\mciteSetBstMidEndSepPunct{\mcitedefaultmidpunct}
{\mcitedefaultendpunct}{\mcitedefaultseppunct}\relax
\EndOfBibitem
\bibitem[Pak \emph{et~al.}(2017)Pak, Lee, Lee, Jang, Ahn, Ma, Cho, Hong, Lee,
  Jeong, Im, Shin, Morris, Cha, Sohn, and Kim]{Pak2017}
S.~Pak, J.~Lee, Y.-W. Lee, A.-R. Jang, S.~Ahn, K.~Y. Ma, Y.~Cho, J.~Hong,
  S.~Lee, H.~Y. Jeong, H.~Im, H.~S. Shin, S.~M. Morris, S.~Cha, J.~I. Sohn and
  J.~M. Kim, \emph{Nano Letters}, 2017, \textbf{17}, 5634--5640\relax
\mciteBstWouldAddEndPuncttrue
\mciteSetBstMidEndSepPunct{\mcitedefaultmidpunct}
{\mcitedefaultendpunct}{\mcitedefaultseppunct}\relax
\EndOfBibitem
\bibitem[Lin \emph{et~al.}(2021)Lin, Ong, Bange, Faria~Junior, Peng, Ziegler,
  Zipfel, B{\"a}uml, Paradiso, Watanabe, Taniguchi, Strunk, Monserrat, Fabian,
  Chernikov, Qiu, Louie, and Lupton]{Lin2021}
K.-Q. Lin, C.~S. Ong, S.~Bange, P.~E. Faria~Junior, B.~Peng, J.~D. Ziegler,
  J.~Zipfel, C.~B{\"a}uml, N.~Paradiso, K.~Watanabe, T.~Taniguchi, C.~Strunk,
  B.~Monserrat, J.~Fabian, A.~Chernikov, D.~Y. Qiu, S.~G. Louie and J.~M.
  Lupton, \emph{Nature Communications}, 2021, \textbf{12}, 5500\relax
\mciteBstWouldAddEndPuncttrue
\mciteSetBstMidEndSepPunct{\mcitedefaultmidpunct}
{\mcitedefaultendpunct}{\mcitedefaultseppunct}\relax
\EndOfBibitem
\bibitem[Zhang and Niu(2015)]{zhang2015chiral}
L.~Zhang and Q.~Niu, \emph{Physical review letters}, 2015, \textbf{115},
  115502\relax
\mciteBstWouldAddEndPuncttrue
\mciteSetBstMidEndSepPunct{\mcitedefaultmidpunct}
{\mcitedefaultendpunct}{\mcitedefaultseppunct}\relax
\EndOfBibitem
\bibitem[Zhang and Murakami(2022)]{Tiantian2022}
T.~Zhang and S.~Murakami, \emph{Phys. Rev. Research}, 2022, \textbf{4},
  L012024\relax
\mciteBstWouldAddEndPuncttrue
\mciteSetBstMidEndSepPunct{\mcitedefaultmidpunct}
{\mcitedefaultendpunct}{\mcitedefaultseppunct}\relax
\EndOfBibitem
\bibitem[Giannozzi \emph{et~al.}(2009)Giannozzi, Baroni, Bonini, Calandra, Car,
  Cavazzoni, Ceresoli, Chiarotti, Cococcioni,
  Dabo,\emph{et~al.}]{giannozzi2009}
P.~Giannozzi, S.~Baroni, N.~Bonini, M.~Calandra, R.~Car, C.~Cavazzoni,
  D.~Ceresoli, G.~L. Chiarotti, M.~Cococcioni, I.~Dabo \emph{et~al.},
  \emph{Journal of physics: Condensed matter}, 2009, \textbf{21}, 395502\relax
\mciteBstWouldAddEndPuncttrue
\mciteSetBstMidEndSepPunct{\mcitedefaultmidpunct}
{\mcitedefaultendpunct}{\mcitedefaultseppunct}\relax
\EndOfBibitem
\bibitem[Schlipf and Gygi(2015)]{Martin2015}
M.~Schlipf and F.~Gygi, \emph{Computer Physics Communications}, 2015,
  \textbf{196}, 36--44\relax
\mciteBstWouldAddEndPuncttrue
\mciteSetBstMidEndSepPunct{\mcitedefaultmidpunct}
{\mcitedefaultendpunct}{\mcitedefaultseppunct}\relax
\EndOfBibitem
\bibitem[Marini \emph{et~al.}(2009)Marini, Hogan, Grüning, and
  Varsano]{Andrea2009}
A.~Marini, C.~Hogan, M.~Grüning and D.~Varsano, \emph{Computer Physics
  Communications}, 2009, \textbf{180}, 1392--1403\relax
\mciteBstWouldAddEndPuncttrue
\mciteSetBstMidEndSepPunct{\mcitedefaultmidpunct}
{\mcitedefaultendpunct}{\mcitedefaultseppunct}\relax
\EndOfBibitem
\bibitem[Sangalli \emph{et~al.}(2019)Sangalli, Ferretti, Miranda, Attaccalite,
  Marri, Cannuccia, Melo, Marsili, Paleari, Marrazzo, Prandini, Bonf{\`{a}},
  Atambo, Affinito, Palummo, Molina-S{\'{a}}nchez, Hogan, Grüning, Varsano,
  and Marini]{Sangalli2019}
D.~Sangalli, A.~Ferretti, H.~Miranda, C.~Attaccalite, I.~Marri, E.~Cannuccia,
  P.~Melo, M.~Marsili, F.~Paleari, A.~Marrazzo, G.~Prandini, P.~Bonf{\`{a}},
  M.~O. Atambo, F.~Affinito, M.~Palummo, A.~Molina-S{\'{a}}nchez, C.~Hogan,
  M.~Grüning, D.~Varsano and A.~Marini, \emph{Journal of Physics: Condensed
  Matter}, 2019, \textbf{31}, 325902\relax
\mciteBstWouldAddEndPuncttrue
\mciteSetBstMidEndSepPunct{\mcitedefaultmidpunct}
{\mcitedefaultendpunct}{\mcitedefaultseppunct}\relax
\EndOfBibitem
\bibitem[Bernardi(2016)]{Bernardi2016}
M.~Bernardi, \emph{The European Physical Journal B}, 2016, \textbf{89},
  239\relax
\mciteBstWouldAddEndPuncttrue
\mciteSetBstMidEndSepPunct{\mcitedefaultmidpunct}
{\mcitedefaultendpunct}{\mcitedefaultseppunct}\relax
\EndOfBibitem
\bibitem[Yu \emph{et~al.}(2021)Yu, Zhou, Wan, and Li]{Yu_2021}
J.~Yu, J.~Zhou, X.~Wan and Q.~Li, \emph{New Journal of Physics}, 2021,
  \textbf{23}, 033005\relax
\mciteBstWouldAddEndPuncttrue
\mciteSetBstMidEndSepPunct{\mcitedefaultmidpunct}
{\mcitedefaultendpunct}{\mcitedefaultseppunct}\relax
\EndOfBibitem
\end{mcitethebibliography}
\bibliographystyle{rsc} } %the RSC's .bst file

\end{document}